\documentclass{aastex}
\usepackage{emulateapj5}
\usepackage{epsfig}
\renewcommand{\placetable}[1]{}
\renewcommand{\placefigure}[1]{}
\renewcommand{\tablecolumns}[1]{}
\renewcommand{\tablewidth}[1]{}
\renewcommand{\tabletypesize}[1]{#1}
\renewcommand{\tablecaption}[1]{\caption{#1}}
\renewcommand{\tablehead}[1]{\hline\hline\noalign{\smallskip} #1}
\renewcommand{\colhead}[1]{#1}
\renewcommand{\startdata}{\\ \hline \noalign{\smallskip}}
\renewcommand{\enddata}{\hline \end{tabular}\smallskip\\}
\renewcommand{\tablenotetext}[2]{\footnotetext{}{\smallskip $^{\rm #1}$#2}\\}
\renewcommand{\tablerefs}[1]{References. --- #1}
\renewcommand{\tablecomments}[1]{Note. --- #1\smallskip\\}
\newcommand{\wcap}{0.476\textwidth}
\submitted{The Astrophysical Journal, 586: \#\#\#--\#\#\#, 2003 April 1, \emph{in press}}

\newcommand{\grays}{$\gamma$-rays}
\newcommand{\Dxx}{D_{xx}}

\newcommand{\e}[2]{$^{#1}_{#2}$}
\newcommand{\fsizeone}{1.1}
\newcommand{\fsizetwo}{2.3} 
\newcommand{\twofigs}{0.49\textwidth}

\newcommand{\adv}{Adv.\ Space Res.}
\newcommand{\plb}{Phys.\ Lett.\ B}
\newcommand{\pub}[4]{#4, #1, #2, #3}
\newcommand{\puba}[4]{#4, #1, #3}

\shorttitle{Challenging cosmic ray propagation with antiprotons}
\shortauthors{Moskalenko et al.}
\begin{document}
\title{Challenging cosmic ray propagation with antiprotons.\\
Evidence for a ``fresh'' nuclei component?}

\author{Igor V.~Moskalenko\altaffilmark{1}}
\affil{NASA/Goddard Space Flight Center, Code 661, Greenbelt, MD 20771}
\altaffiltext{1}{Joint Center for Astrophysics, University of Maryland, 
   Baltimore County, Baltimore, MD 21250}
\email{igor.moskalenko@gsfc.nasa.gov}

\author{Andrew W.~Strong}
\affil{Max-Planck-Institut f\"ur extraterrestrische Physik,
   Postfach 1603, D-85740 Garching, Germany}
\email{aws@mpe.mpg.de}

\author{Stepan G.~Mashnik}
\affil{Los Alamos National Laboratory, Los Alamos, NM 97545}
\email{mashnik@lanl.gov}

\and

\author{Jonathan F.~Ormes} 
\affil{NASA/Goddard Space Flight Center, Code 600, Greenbelt, MD 20771}
\email{Jonathan.F.Ormes@gsfc.nasa.gov}


\begin{abstract}
Recent measurements of the cosmic ray (CR) antiproton flux
have been shown to challenge existing CR propagation
models. 
It was shown that the reacceleration models designed to match
secondary to primary nuclei ratios (e.g., boron/carbon) produce too few
antiprotons. Matching
both the secondary to primary nuclei ratio and the antiproton flux requires
artificial breaks in the diffusion coefficient and the primary
injection spectrum suggesting the need for other approaches.

In the present paper we discuss one possibility to overcome these
difficulties. Using the
measured antiproton flux \emph{and}
B/C ratio to fix the diffusion coefficient, we show that 
the spectra of primary nuclei as measured in the heliosphere may contain a
fresh local ``unprocessed'' component at low energies
perhaps associated with the Local Bubble, thus decreasing
the measured secondary to primary nuclei ratio.
The independent evidence for SN activity in the solar vicinity
in the last few Myr supports this idea.
The model reproduces
antiprotons, B/C ratio, and elemental abundances up to Ni ($Z\leq28$).
Calculated isotopic distributions of Be and B 
are in perfect agreement with CR data.
The abundances of three ``radioactive clock'' isotopes in CR, \e{10}{}Be,
\e{26}{}Al, \e{36}{}Cl,
are all consistent and indicate a halo size $z_h\sim4$ kpc based on 
the most accurate data taken by the ACE spacecraft.

\end{abstract}



\section{INTRODUCTION} \label{sec:intro}
The spectrum and origin of antiprotons in CR has been a matter of active
debate since the first reported detections in balloon flights 
\citep{Golden79,bogomolov1}.  Because of the baryonic
asymmetry of the Universe, antiprotons are not found at rest. 
There is a consensus that most of
the CR antiprotons observed near the Earth are ``secondaries'' produced in
collisions of energetic CR particles with interstellar gas
\citep[e.g.,][]{mitchell}.

The spectrum of secondary antiprotons has a peak at about 2 GeV
decreasing sharply towards lower energies. This unique shape
distinguishes antiprotons from other cosmic-ray species and allows for
searches of primary antiprotons at low energies.  Over the last few years
the accuracy has been improved sufficiently
\citep[BESS 1995--2000,][]{Orito00,Sanu00,Asaoka02} that we can 
restrict the spectrum of the secondary component accurately enough to
test Galactic CR propagation models, and the heliospheric modulation.

It has been recently shown \citep{M02} that accurate antiproton
measurements during the last solar minimum 1995--1997
\citep[BESS,][]{Orito00} are inconsistent with existing propagation models
at the $\sim40$\% level at about 2 GeV while the stated measurement 
uncertainties in this energy range are now $\sim20$\%.
The conventional models based on local CR measurements,
simple energy dependence of the diffusion coefficient,  and uniform CR
source spectra throughout the Galaxy fail to reproduce
simultaneously both the secondary to primary nuclei ratio
and antiproton flux.

The reacceleration model designed to match secondary to primary nuclei
ratios (e.g., boron/carbon) produce too few antiprotons because, e.g.,
matching the B/C ratio at all energies requires the diffusion
coefficient to be too large.  The models without reacceleration can
reproduce the antiproton flux, however they fall short of explaining the
low-energy decrease in the secondary to primary nuclei ratio.  To be
consistent with both, the introduction of breaks in the
diffusion coefficient and the injection spectrum is required, which
would suggest new phenomena in particle acceleration and propagation.

Recently there has appeared some indication that the atmospheric contribution
to the antiproton flux measured in the upper atmosphere is underestimated.
If this is true, the reacceleration model could still be the best one to
describe propagation of nucleon species in the Galaxy (for more
details see Section~\ref{sec:conclusion}). 
However, in this work we have assumed the published 
Galactic antiproton flux, corrected for atmospheric 
production, is accurate.

In the present paper we discuss another possibility to overcome the
difficulties encountered by reacceleration models. 
We will show that the inclusion of a local primary component 
at low energies, perhaps associated with the Local Bubble (LB),
reconciles the data.

\smallskip
\section{INTERSTELLAR COSMIC RAY SPECTRUM} \label{sec:spectrum}
Just as secondary nuclei, the product of the disintegration of primary nuclei,
are abundant in CR but rare in the interstellar medium (ISM),
diffuse \grays, antiprotons, and positrons are secondary
products of interactions of mostly CR protons and helium nuclei with
interstellar gas. The CR propagation model that describes any
secondary to primary ratio should equally well describe all the others: B/C,
sub-Fe/Fe, $\bar p/p$ ratios, and spectra of nuclei, positrons, and
diffuse Galactic continuum \grays.

The diffusive reacceleration models naturally reproduce
secondary to primary nuclei ratios in CR and agree better with
K-capture parent/daughter nuclei ratio \citep[e.g., see][]{jones01},
though this result is not completely conclusive due to the large error
bars in CR measurements and uncertainties in important isotopic
cross sections. It is, however, clear that some reacceleration is
unavoidable in the ISM.
Because of the unique shape of the antiproton
spectrum, reacceleration has much weaker effect on it than in
the case of nuclei. Taking into account that the
antiproton production spectrum can be calculated
accurately, antiprotons provide a unique complementary
tool to test propagation models
(and heliospheric modulation).

Our previous result \citep{M01,M02}, in agreement with calculations of
other authors \citep{simon}, was that matching the
secondary/primary nuclei ratio B/C using reacceleration models leads
to values of the spatial diffusion coefficient apparently too large to produce
the required antiproton flux, when the propagated nucleon spectra are
tuned to match the local proton and helium flux measurements. This is an
essential shortcoming.  

Assuming the measured antiproton flux is correct and the 
current heliospheric modulation models are approximately
right, we have the following possibility to reconcile the B/C ratio with the
required flux of secondary antiprotons.  The spectra of
primary nuclei as measured in the heliosphere may contain a fresh
local ``unprocessed'' component at low energies thus decreasing the
measured secondary to primary nuclei ratio.  
This component would have to be local in the sense of being 
specific to the solar neighbourhood, so that the well-known ``Local Bubble'' 
phenomenon is a natural candidate.

The idea that CR are accelerated out
of SN ejecta-enriched matter in  superbubbles has been discussed
in numerous papers \citep[e.g.,][and references therein]{higdon}. 
The possibility that the ``fresh'' component is coming
from the Local Bubble has been discussed by, e.g.,
\citet{morfill} and \citet{davis}. We will hereafter call it the
``Local Bubble Hypothesis.'' 
The idea is that primary CR like $^{12}$C and  $^{16}$O have a local component at 
low energies, while secondary CR like B are produced Galaxy-wide over 
the confinement time of 10--100 Myr. 
Then the B/C ratio will be lower at low energies than expected 
in a uniform model, due to the enhanced local C (and the reduced 
Galactic production of B).
If this idea is correct then the high-energy
part of the secondary/primary nuclei ratio plus the measured antiproton
flux at maximum, $\sim 2$ GeV, can be used to restrict the value and
energy dependence of the diffusion coefficient,
while the required contribution of the local sources can be
derived from the measured secondary/primary nuclei ratio at low
energies. 

One additional hint for the possible existence of an ``unprocessed''
component is the calculated ratio of $^{13}$C/$^{12}$C $\sim0.14$ at 120
MeV/nucleon (modulation potential 500 MV), which appears to be a
factor of two larger than that observed when the
propagation parameters are tuned to the B/C ratio \citep{M02}. The
isotope $^{13}$C is almost all secondary, as are Be and B
isotopes.  Since the primary source of $^{13}$C is $^{16}$O, accounting
for as much as $\sim60$\% of the total, this may indicate an
``over-enrichment'' of the assumed source abundances in oxygen.  If
so, the ``over-enrichment'' may be true also for primary carbon,  but
tuning to the observed B/C artificially  eliminates the excess of
lithium, beryllium, and boron.   We note that the ratio of
$^{15}$N/$^{16}$O is, however, correct, and the problem with
overproduction of $^{13}$C may arise from  cross section errors
\footnote{The production cross sections of $^{14,15}$N, and
$^{12,13}$C have been measured only in a narrow energy range.}
(see Appendix~\ref{cs}).

\section{THE LOCAL BUBBLE HYPOTHESIS}
The low-density region around the sun, filled with hot H\ {\sc i} gas,
is called the Local Bubble \citep[e.g.,][]{sfeir}. 
The size of the region is about 200 pc, 
and it is likely that it was produced in a series of supernova
(SN) explosions.  Most probably its progenitor was an OB 
star association. Though people discuss different scenarios
\citep[e.g.,][]{maiz,berghofer},  the LB age and the number of SN
progenitors appears to be similar, $\sim10$ Myr and $\sim 10-20$ SN,
respectively.  Most probably they exploded as core-collapse SN II
or thermonuclear SN Ib/c with a mass of pre-SN stars between several 
and $\sim10 M_\sun$, with the last SN explosion occuring approximately 
1--2 Myr ago, or 3 SN occuring  within the last 5 Myr. 

There is also some evidence of a SN explosion nearby.  An excess of
$^{60}$Fe measured in a deep ocean core sample of
ferromanganese crust suggests the
deposition of SN-produced iron on earth \citep{fe60}.  The enhanced
concentrations were found in two of three layers corresponding to a
time span of $<2.8$ Myr and 3.7--5.9 Myr, respectively. The study
suggests a SN explosion about 5 Myr ago at 30 pc distance. Another
study reports an enhancement in the CR intensity dated about 40 kyr
ago \citep{sonett}, which is interpreted as the passage 
across the solar system of the shock
wave from a SN exploding about 0.1 Myr ago.
Taking into account possible errors of all these estimates, they
point to a nearby SN explosion some 1 Myr ago
\citep*[see also discussion in][]{benitez}.

It could also be that ``fresh'' LB contributions from
continuous acceleration in the form of shock waves \citep{bykov},
and/or energetic particles coming directly from SN remnants still 
influence the spectra and
abundances of local CR.
The elemental abundances of the low-energy nonthermal component
in a superbubble
can differ strongly from the standard cosmic abundances 
\citep{bykov2} due to
ejection of matter enriched with heavy elements from SN and 
stellar winds of massive stars (Wolf-Rayet, OB stars).
The continuous acceleration is connected with the lifetime of a shock
wave in the LB. A reasonable estimate is given by 
the sound crossing time, approximately 2 Myr, for a distance of 200 pc 
in a $10^6$ K plasma \citep{berghofer}.
On the other hand, the particle crossing time can be estimated as 
$t\sim x^2/D\sim1$ Myr for a typical value of the diffusion
coefficient in the ISM $D\sim10^{28}$ cm s$^{-2}$ and $x\sim200$ pc.
Therefore, accelerated particles are expected 
to be present in this region.

\placetable{table1}


\begin{table*}[!bh]
\vspace{-1\baselineskip}
\tablecolumns{7}
\tablewidth{0mm}
\tablecaption{Propagation parameter sets.
\label{table1}}

\begin{tabular}{lcccccc}

\tablehead{
\colhead{} & 
\colhead{} & 
\multicolumn{2}{c}{Diffusion coefficient\tablenotemark{a}} & 
\colhead{} &
\colhead{Alfv\'en speed,\tablenotemark{b}} &
\colhead{Source}
\\
\cline{3-4}
\colhead{Model} & 
\colhead{Injection index, $\gamma$} &
\colhead{$D_0$, cm$^2$ s$^{-1}$} & 
\colhead{Index, $\delta$} &
\colhead{} &
\colhead{$v_A/\sqrt w$, km s$^{-1}$} &
\colhead{abundances}
}
\startdata
Plain Diffusion (PD) &
2.16                 &
$3.10\times10^{28}$  & 
0.60                 &
 &
---                  &
---
\smallskip\\
 
Diffusive            \\
Reacceleration I (DR I)  &
2.28                 &
$3.30\times10^{28}$  & 
0.47                 &
 & 
23                   &
LBS=CRS              
\smallskip\\

Diffusive            \\
Reacceleration II (DR II)  &
2.28                 &
$3.30\times10^{28}$  & 
0.47                 &
 & 
23                   &
LBS$\neq$CRS         
\\

\enddata
\footnotesize
\tablecomments{Adopted halo size $z_h=4$ kpc.\vspace{-0.3\baselineskip}}
\tablenotetext{a}{$\rho_0=3$ GV.\vspace{-0.3\baselineskip}}
\tablenotetext{b}{$v_A$ is the Alfv\'en speed, and
$w$ is defined as the ratio of MHD wave energy density to magnetic field energy density.}
\end{table*}

\section{THE CALCULATION PROCEDURE}
In our calculations we use the propagation model 
GALPROP\footnote{GALPROP model including software and data sets is 
available at \url{http://www.gamma.mpe-garching.mpg.de/$\sim$aws/aws.html}}
as described elsewhere \citep{SM98,M02}; for the present purpose 
the 2D cylindrically symmetrical option is sufficient. 
For a given halo height $z_h$ the diffusion coefficient as a function of 
momentum
and the reacceleration or convection parameters is determined by data
on  secondary-to-primary ratios in CR. The spatial diffusion coefficient is
taken as $\Dxx = \beta D_0(\rho/\rho_0)^{\delta}$;  
the corresponding diffusion in momentum space and other details of the 
models can be found in our earlier papers.
We use our standard methodologies and include two types
of cosmic ray sources.

Antiproton production and propagation are calculated as described
in \citet{M02}.

The nucleon injection spectrum of the
Galactic component was taken as a modified power-law in rigidity
\citep{jones}, $dq(p)/d\rho \propto
\rho^{-\gamma}/\sqrt{1+(\rho/2)^{-2}}$, for the injected particle density.
The proton and He spectra are tuned to the local measurements as described
in \citet{M02}.

The LB spectrum is taken to have the form (as suggested by 
\citealt{bykov} and \citealt{bykov2}
for continous acceleration by interstellar shocks):
$df/d\rho \propto \rho^{-\eta}\exp(-\rho/\rho_b)$, where $\rho$ is the
rigidity, and $\rho_b$ is the cut off rigidity. $\rho_b$ and the LB 
source abundances are adjustable parameters.
In terms of kinetic energy per nucleon $E$ this can be re-written as
\begin{equation}
\label{eq1}
\frac{df}{dE}=a(Z,A)\frac{A(E+m)}{Zp}\rho^{-\eta}\exp(-\rho/\rho_b),
\end{equation}
where $a(Z,A)$ is the abundance of a nucleus $(Z,A)$, $Z,A$ are the nucleus
charge and atomic number, correspondingly, $m$ is the atomic mass unit,
$p$ is the momentum per nucleon, $\rho_b=\frac{A}{Z}\sqrt{(E_b +m)^2 -m^2}$.
The particular spectral shape of the LB
component is not important as long as it decreases sharply towards high
energies and is much softer than the Galactic CR spectrum.  
We show the results obtained with $\eta=1$, but they are
very similar  with $\eta=2$ with $\rho_b$ adjusted correspondingly.

The procedure to tune the CR elemental abundances, secondary/primary
nuclei ratios, and antiproton flux we adopted was as follows.
The high energy part of B/C ratio and antiproton flux measurements
are used to restrict the value of the diffusion coefficient and its
energy dependence, while the low energy part of B/C ratio is used
to fix a value for the reacceleration level and define 
the parameters of the LB component.

The heliospheric modulation is
treated using the force-field approximation \citep{forcefield}.
In our previous paper \citep{M02} we used the
best currently available model of the heliospheric modulation, 
the steady-state drift model. The use of the simpler model
in the current paper is justified by the following arguments.
First, in the previous paper \citep{M02} we tried to build
a model of particle propagation explaining both CR nuclei and 
antiproton fluxes equally well. The heliospheric modulation
calculation was considered as one of the main possible reasons for
the discrepancy between the nuclei and antiproton fluxes.
It appears however that the interstellar propagation and/or
cross section errors are responsible for the nuclei/antiproton discrepancy,
unless the current models of heliospheric modulation are 
\emph{completely} wrong, which is not very likely.
Second, the current paper evaluates a new hypothesis for the
origin of CR where the main emphasis is  on the local
CR component. Use of a sophisticated modulation model
containing additional unknown variables will 
make such evaluation unnecessarily complicated. 
Third, the parameters of the diffusion coefficient are fixed
using the data above few GeV where the modulation is weak.

We consider three different models
(Table \ref{table1}) with parameters 
fixed using the described procedure.
They are: the simplest plain
diffusion model (PD), and two reacceleration models, which use different
assumed isotopic abundances in the Galactic CR sources and LB
sources. Diffusive reacceleration model I (DR I) has the same
isotopic abundances in Galactic CR and LB sources.
Diffusive reacceleration model II (DR II) is the same as DR I except that
the LB isotopic abundances are tuned to
match the low-energy data by ACE and Ulysses, thus increasing the
freedom to fit the data.
The Galactic CR elemental source abundances (DR II model) are tuned
to the abundances measured at high
energies where the heliospheric modulation is weak.

\begin{figure*}[!th] 
\epsscale{\fsizetwo}
\plottwo{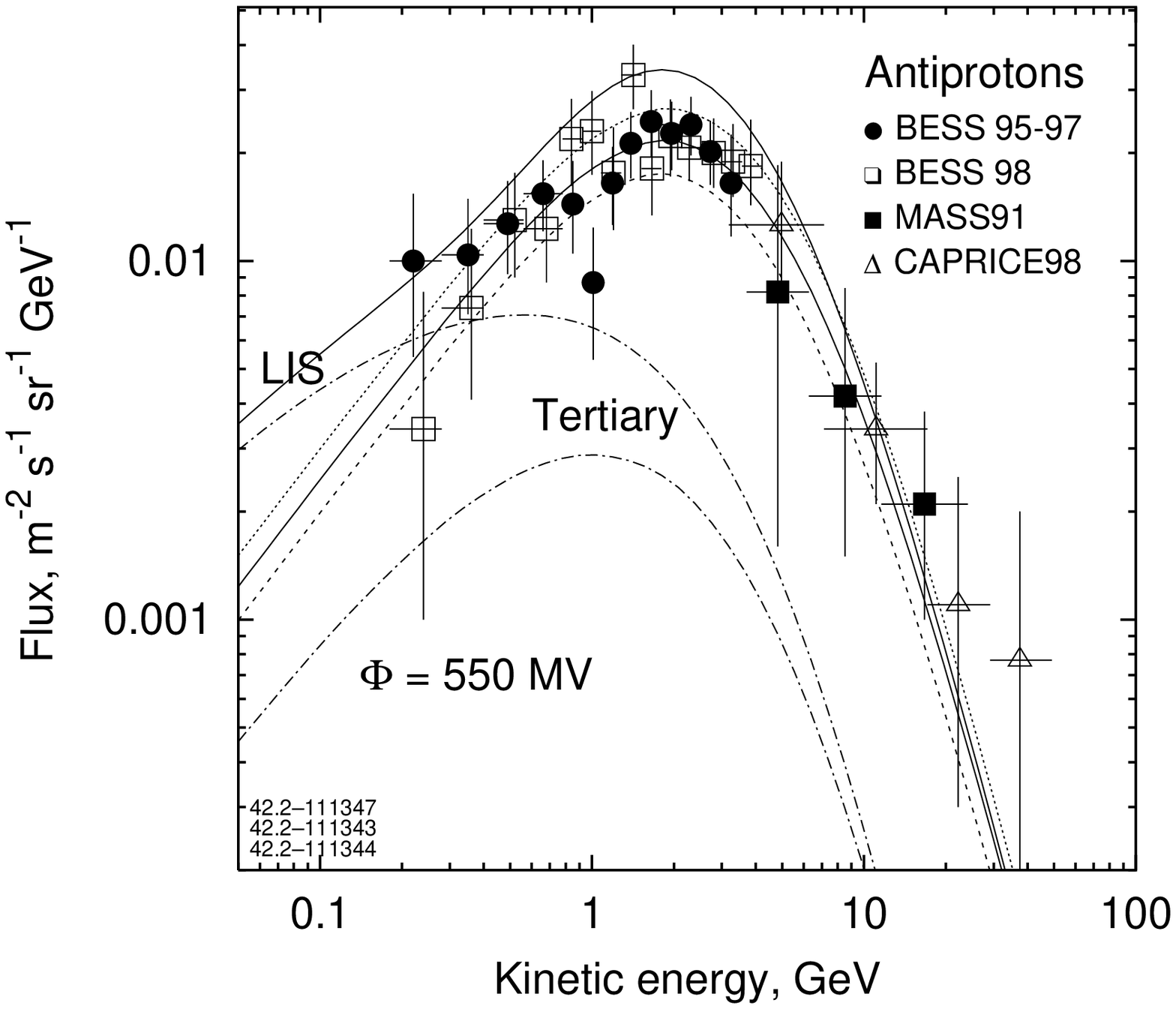}{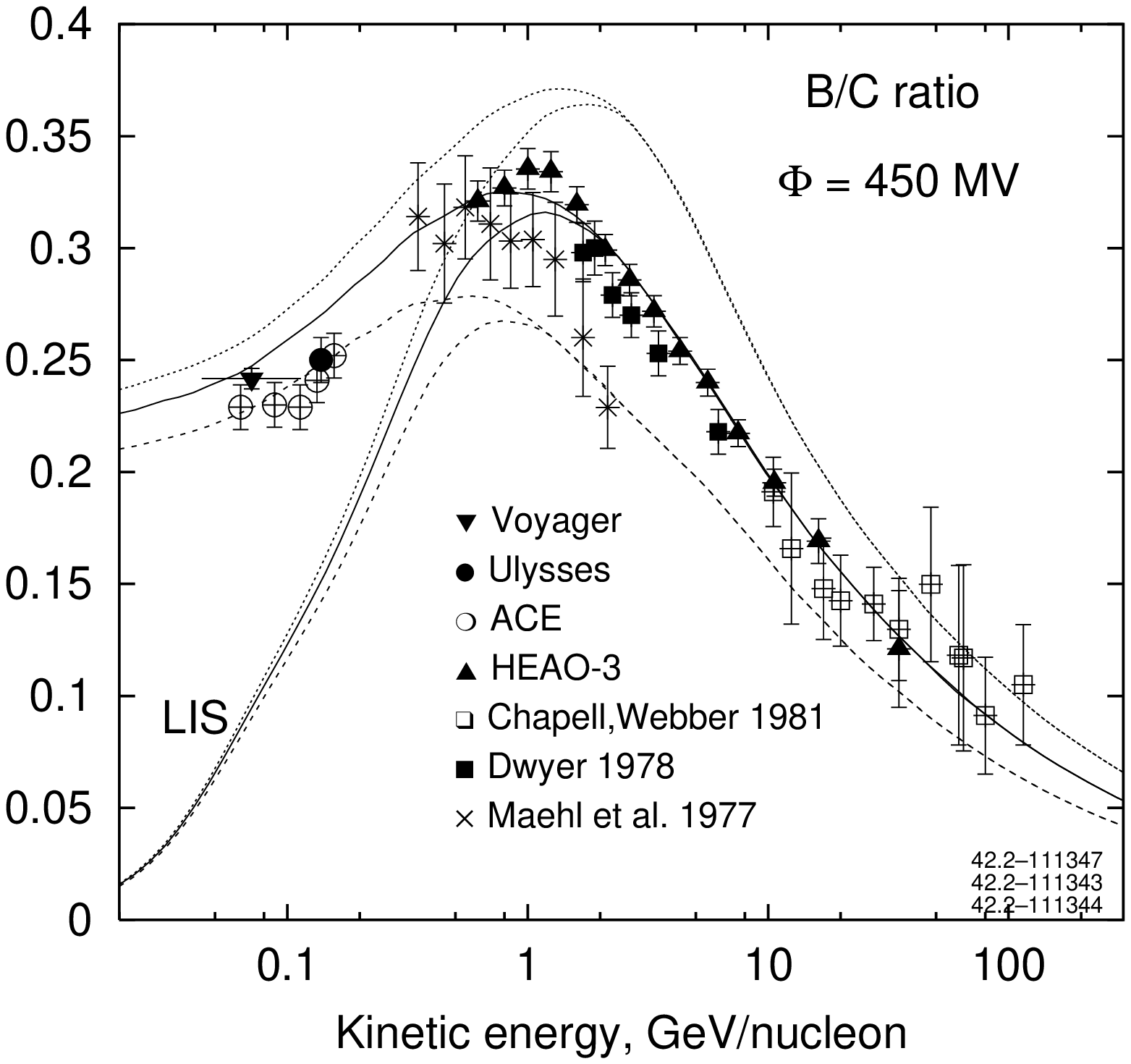}
\caption[f1a.eps,f1b.eps]{
\emph{Left:} Antiproton flux calculated in DR I/II models 
with index $\delta=0.47$ in the diffusion coefficient and different normalization
values, $D_0$ cm s$^{-2}$.
Solid curves -- $D_0=3.3\times10^{28}$ at $\rho_0=3$ GV, 
upper curve -- local interstellar (LIS), lower curve -- modulated. Dots --  
$D_0=2.6\times10^{28}$, shown modulated only, dashes -- 
$D_0=4.3\times10^{28}$, shown modulated only. 
The two lowest
curves (dash-dot) marked ``tertiary'' show separately the LIS spectrum and
modulated ``tertiary'' component for $D_0=3.3\times10^{28}$ cm s$^{-2}$.
Modulation was made with $\Phi=550$ MV 
(force field).
Data: BESS 95-97 \citep{Orito00}, BESS 98 \citep{Asaoka02}
MASS91 \citep{Stoc01}, CAPRICE98 \citep{boezio01}.
\emph{Right:} B/C ratio calculated with LB contribution,
$E_b=500$ MeV. The lines are coded as on the left panel.
Lower curves -- LIS, upper -- modulated ($\Phi=450$ MV).
Data below 200 MeV/nucleon: ACE \citep{davis}, Ulysses
\citep*{ulysses_bc}, Voyager \citep*{voyager}; high energy data:
HEAO-3 \citep{Engelmann90}, for other references see
\citet{StephensStreitmatter98}.
\label{fig:pbars}}
\vspace{-1\baselineskip}
\end{figure*}

\placefigure{fig:pbars}

The plain diffusion model, without an LB component, has already been
discussed in \citet{M02}. It is inconsistent with 
low energy data on secondary/primary  ratios, and
at high energies matching the B/C ratio would cause an overproduction
of antiprotons.
We do not see a plausible modification of this model, even including an LB component,
which would allow to us simultaneously fit antiprotons and the B/C ratio.

Hence we turn to the models with reacceleration.
Fig.~\ref{fig:pbars} (left) illustrates the process of fixing the
normalization of the diffusion coefficient using antiprotons. 
The antiproton flux shown as calculated in the DR I/II models with
$\delta=0.47$ and different normalizations in the diffusion coefficient,
$D_0=2.6\times10^{28}, 3.3\times10^{28}, 4.3\times10^{28}$ 
cm s$^{-2}$ at $\rho=3$ GV (for antiprotons this corresponds to 
kinetic energy $\sim2$ GeV). The injection index $\gamma$ is taken
equal to 2.28, and the Alfv\'en speed $v_A=23$ km s$^{-1}$.
The antiproton flux at maximum, $\sim2$ GeV, appears
to be quite sensitive to the value of the diffusion coefficient and
allows us to fix it at $D_0=(3.3\pm0.8)\times10^{28}$ cm s$^{-2}$.
(A 1$\sigma$ deviation in the data translates to approximately
$-20$\%$+25$\% accuracy in $D_0$.)
The exact value of $\delta$ is not critical since we compare with
the antiproton measurements at maximum, $\sim2$ GeV.
Inelastically scattered antiprotons, the ``tertiary'' component,
appears to be important at low energies only in the ISM.
Fig.~\ref{fig:pbars} (right) shows corresponding calculations of the B/C ratio.
A halo height of $z_h=4$ kpc is used \citep{SM01,MMS}; 
using values differing by up to a factor 2 
(the estimated uncertainty) would not affect the 
conclusions, 
the diffusion coefficient would simply scale accordingly, $D_0\propto z_h$
approximately.
In Section \ref{sec:results}.2 we re-evaluate the radioactive secondaries
for the current model and
show that our adopted value of $z_h$ is in good agreement
with the data.

The local interstellar (LIS) proton and helium spectra used in our 
antiproton calculations are the best fits
to the data as described in \citet{M02}. The proton spectrum
is shown in Fig.~\ref{fig:protons} together with data.
Because of the measurements with large statistics, mostly by BESS
\citep{Sanu00} and AMS \citep{p_ams}, and weak 
heliospheric modulation above 10 GeV, the error arising from 
uncertainties in the primary spectra
is only $\sim5$\%. 
The agreement between BESS and AMS data, currently considered to be the
most accurate, is impressive. The data collected by other instruments
(IMAX, CAPRICE, LEAP) at approximately the same solar modulation level are 
lower by 5--10\% and have larger error bars.
Adopting a smaller LIS proton flux would lead to an even more
dramatic discrepancy between the antiproton flux data and the calculation in the
``standard'' reacceleration model \citep{M02}. 
In any case, allowing for a $\sim5-10$\% 
systematic uncertainty in the proton measurements does not change the 
conclusions of the present paper because the antiproton data have
larger error bars.

We note that as in the case of other nuclei, there
should be an LB contribution to proton and He spectra.
The Galactic injection
spectra of protons and He should thus be significantly flatter 
below several GeV to match the data points at low energies.
This does not influence the antiproton production because
(i) the LB does not produce a significant amount of secondaries, and (ii)
the antiproton threshold production energy is high, $\sim10$ GeV.

Further tuning can be done
using the high energy part of the B/C ratio, which is not influenced
by heliospheric modulation and supposedly contains only a Galactic
component of CR. Fig.~\ref{fig:BC300} (left) shows a calculation of the B/C ratio
for $E_b=500$ MeV and different energy dependencies of
the diffusion coefficient.  The plotted curves correspond to values
of the power-law index $\delta=0.42, 0.47, 0.52$, while the injection index was
tuned to match the high energy spectral data. Index $\delta\sim0.47$
is chosen as giving the best match. A pure Kolmogorov spectrum,
$\delta=1/3$, thus seems to be excluded by antiproton spectrum data 
taken in combination with B/C ratio data.

The B/C ratio as calculated with and without a contribution of the LB
component is shown in Fig.~\ref{fig:BC300} (right). 
The LB component is shown
calculated with $E_b=400$, 500, 600 MeV. It is seen,
however, that all three provide 
%
\centerline{\psfig{file=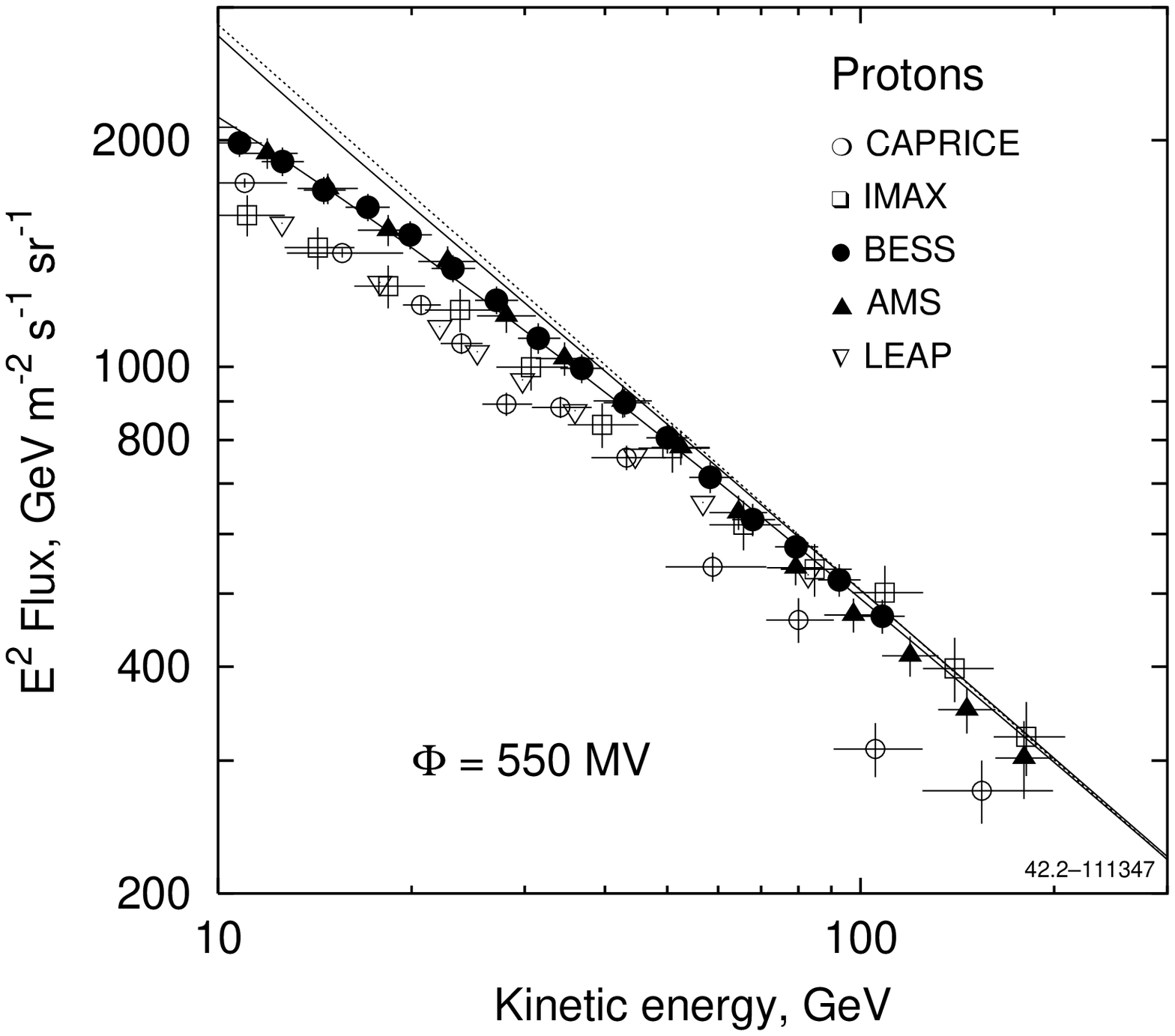,width=\twofigs,clip=}}
\figcaption[f2.eps]
{The proton spectrum as calculated in models DR I/II  
compared with the data (upper curve - LIS, lower - modulated to 550 MV).
Dotted line shows the LIS spectrum best fitted to the data above 20 GeV \citep{M02}. 
Data: IMAX \citep{Menn00}, CAPRICE \citep{Boez99}, BESS \citep{Sanu00}, 
AMS \citep{p_ams}, LEAP \citep{p_leap}. \vspace{1\baselineskip}
\label{fig:protons}}
\noindent 
good agreement with B/C data.
Fig.~\ref{fig:C} shows the interstellar and LB carbon spectra in this model.
By including the LB component, we have therefore been able to obtain a model
simultaneously fitting $p$, He, $\bar p$ and B/C data.
It now remains to apply this model to the full range of CR isotopes.

\subsection{Calculation uncertainties}

We do not discuss here possible calculation \emph{errors}.
Derivation of such errors is a \emph{very} complicated matter
given the many uncertainties in the
input, such as the cross sections, gas distribution in the Galaxy,
systematic errors in the CR measurements, heliospheric modulation,
atmospheric corrections etc. Some possible errors and their effects
have been discussed in \citet{M02}. Here we qualitatively mention
what we think may affect our conclusions and what may not.

Possible errors in the cross
section of the abundant CR nuclei seem to be less important as
they can be compensated by relatively small corresponding
adjustments in the primary abundances. They however may be more
important in case of less abundant ``secondary'' nuclei.
The cross section errors are extensively discussed throughout
this paper. 

Errors in the Galactic gas distribution are not so important in the
case of stable and long-lived nuclei. Such errors are compensated
simultaneously for all species
by the corresponding adjustment of the propagation parameters
(diffusion coefficient). 

Heliospheric modulation may introduce some error, but it will
be similar for all CR nuclei (or their ratios) because the
critical parameter here is the charge-to-mass ratio, approximately
1/2 for all nuclei except (anti-)protons.

\begin{figure*}[!th] 
\epsscale{\fsizetwo}
\plottwo{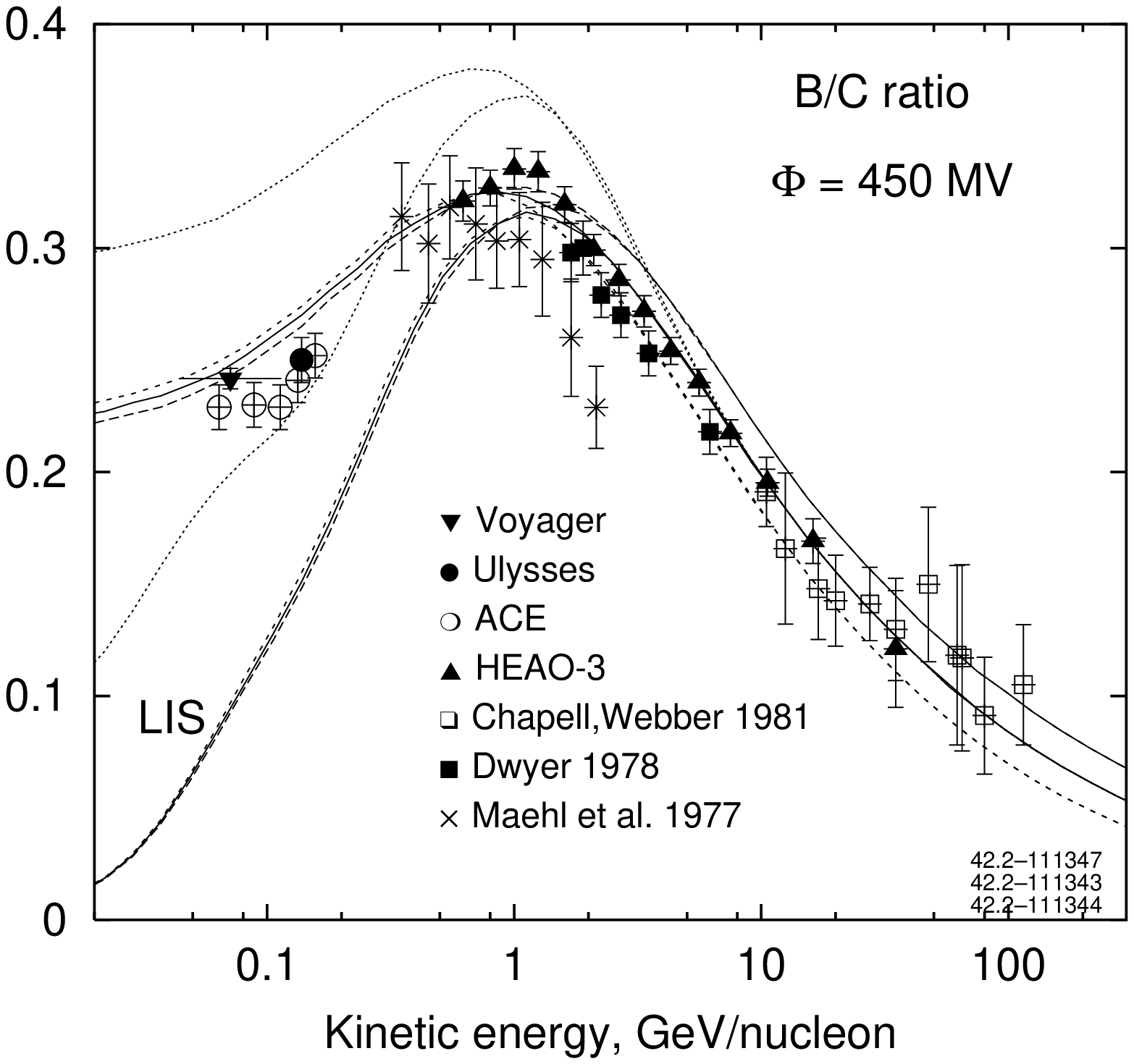}{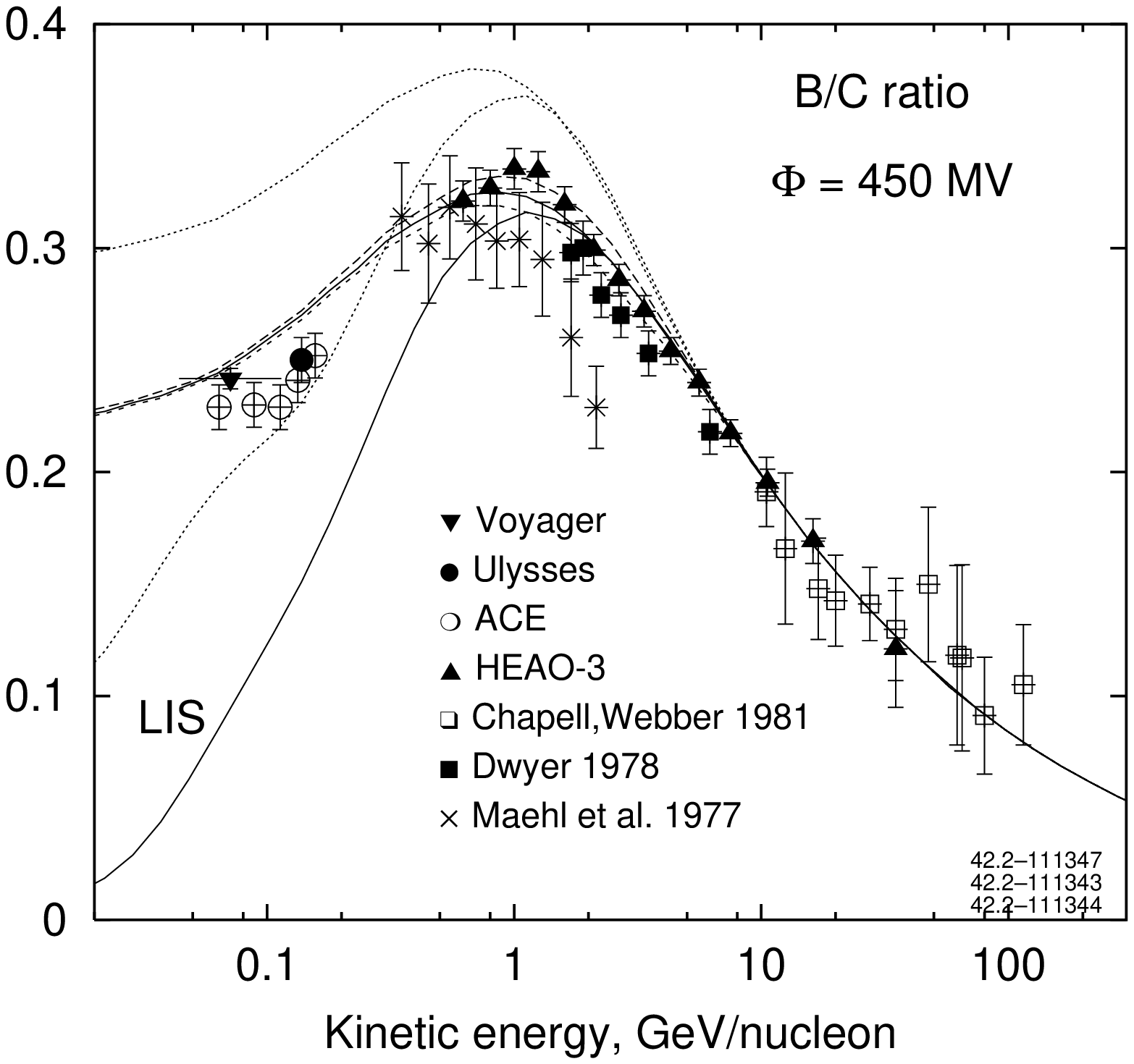}
\caption[f3a.eps,f3b.eps]{ 
\emph{Left:} B/C ratio calculated without (dots) and with
(solid, dashes) LB contribution (DR I/II models),
$E_b=500$ MeV/nucleon, with different energy dependence in
the diffusion coefficient, $\delta=0.52$ (short dashes), 0.47 (solid),
0.42 (dashes). Lower curves -- interstellar (LIS), upper -- modulated 
($\Phi=450$ MV). Data as in Fig.~\ref{fig:pbars} (right).
\emph{Right:} B/C ratio calculated without LB contribution (dotted),
and with $E_b=400$ (long dash), 500 (solid), 600 MeV (short dash).
Lower curves LIS, upper modulated ($\Phi=450$ MV). 
Data as in Fig.~\ref{fig:pbars} (right). \vspace{-1\baselineskip}
\label{fig:BC300} }
\end{figure*}

\placefigure{fig:BC300}

Systematic measurement errors are difficult to account for,
but their effect can be reduced by careful choice of the data to
rely on. This is what we try to do in the present paper.

\placefigure{fig:C}
%
\centerline{\psfig{file=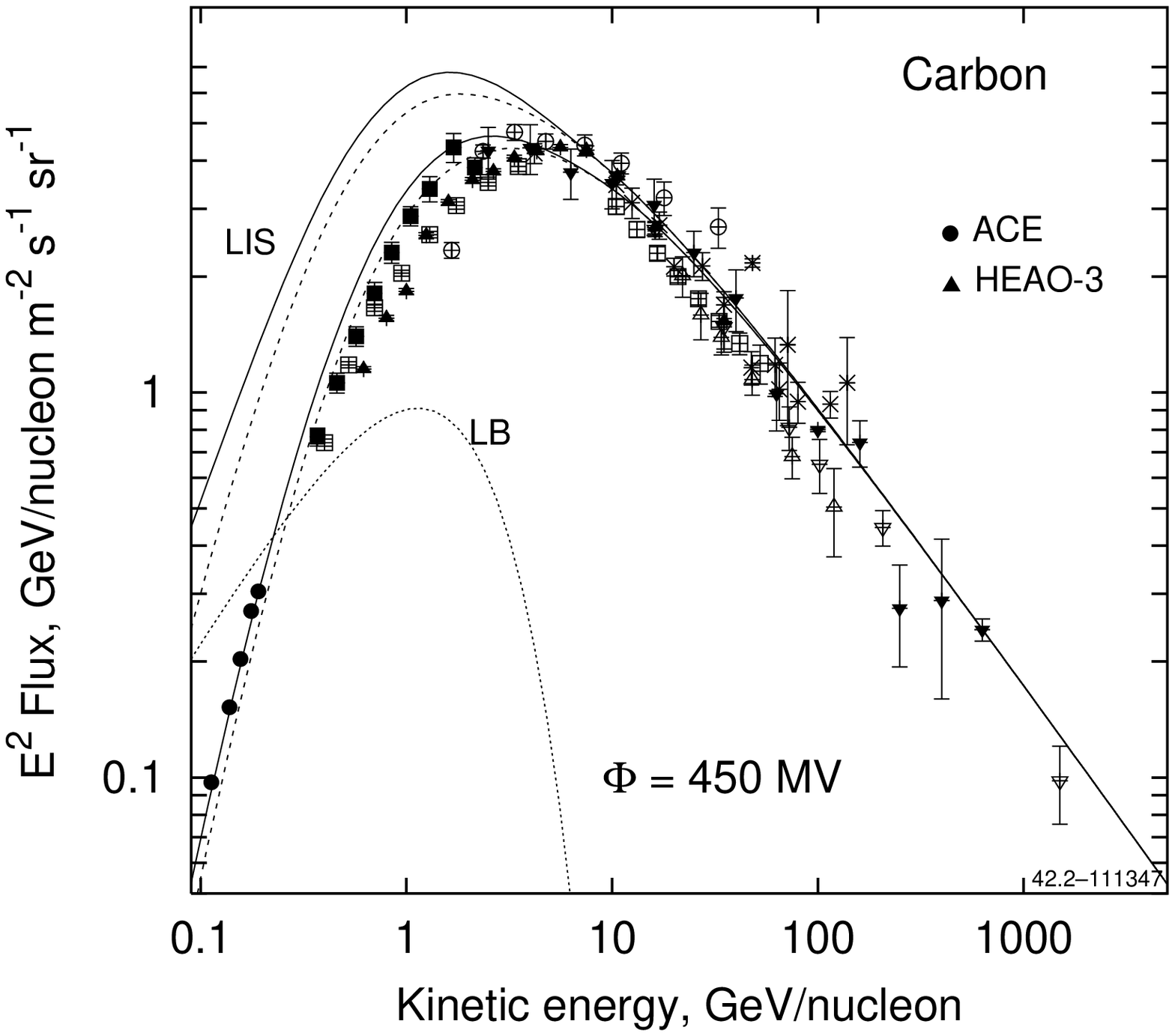,width=\twofigs,clip=}}
\figcaption[f4.eps]{Spectrum of carbon calculated with (solid line, model DR II)
and without (dashes) LB contribution.
Upper curves - LIS, lower curves modulated using force field
approximation ($\Phi=450$ MV). 
Local Bubble (LB) interstellar spectrum is shown by dots.
Data as in Fig.~\ref{fig:B-Fe}. \vspace{2\baselineskip}
\label{fig:C}}
\noindent

\begin{figure*}[!th] 
\psfig{file=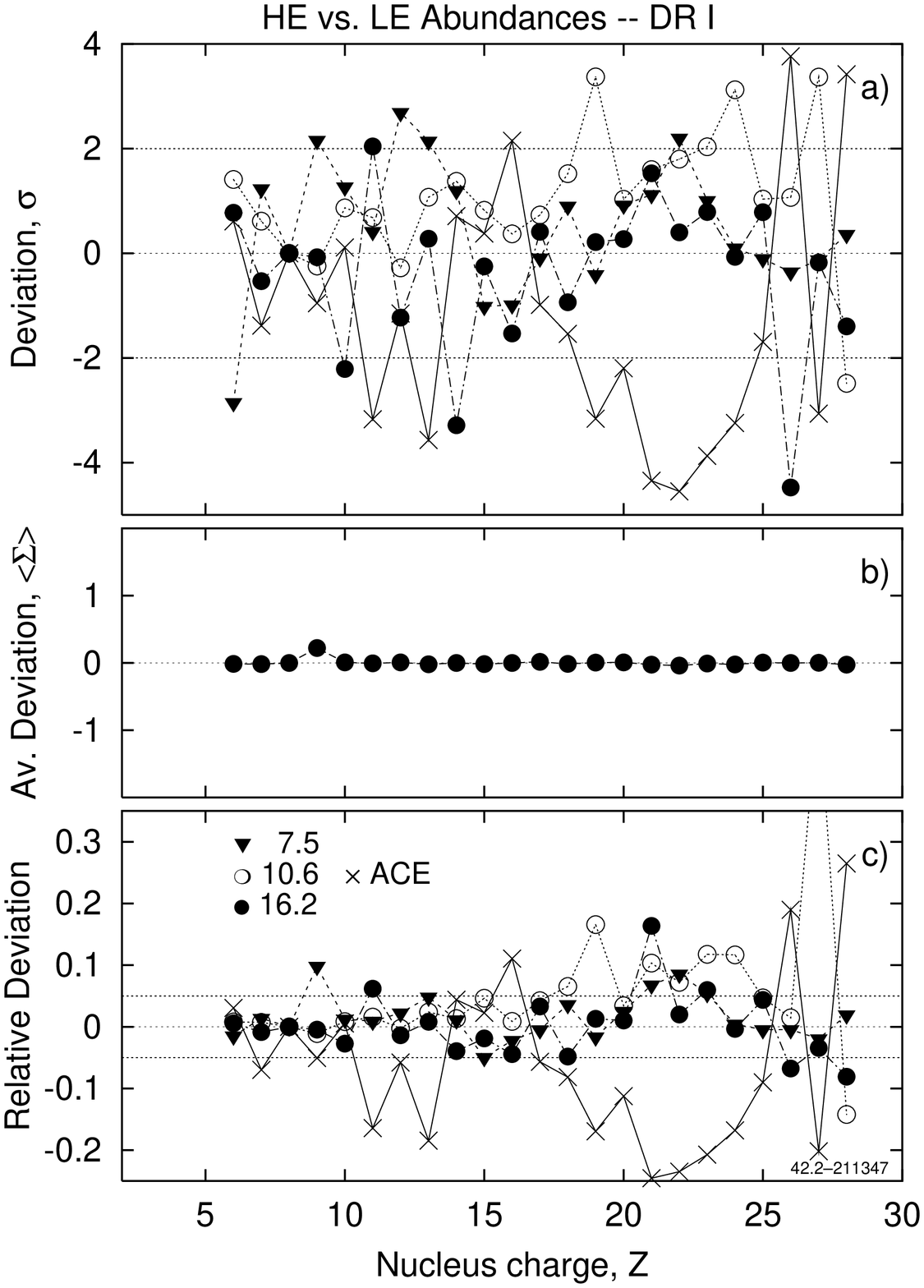,width=\twofigs,clip=}
\psfig{file=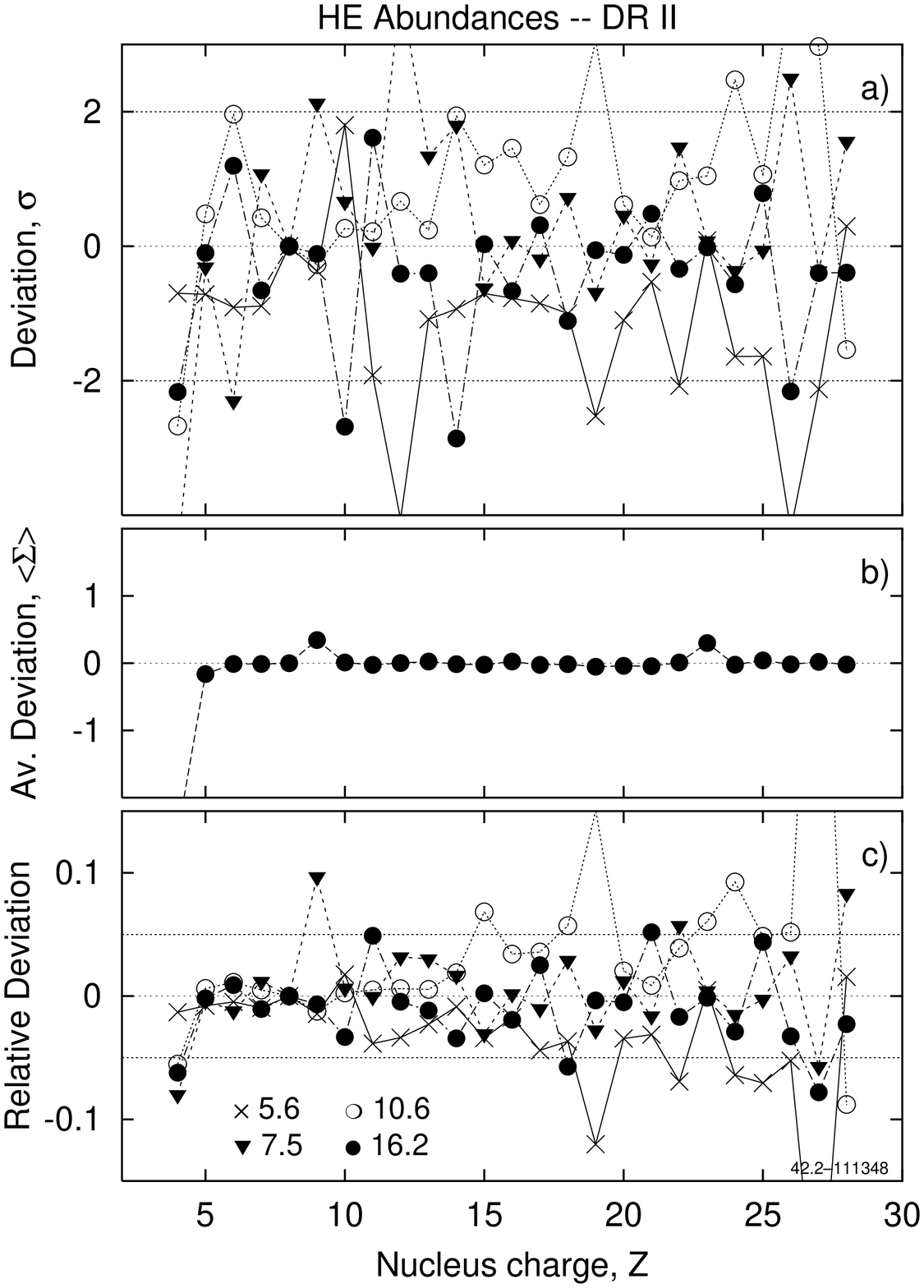,width=\twofigs,clip=}
\begin{minipage}[tl]{\wcap}
\figcaption[f5.eps]{Deviation of propagated abundances (model DR I) from measured 
by HEAO-3 at 7.5, 10.6, 16.2 GeV/nucleon \citep{Engelmann90} 
and ACE at 200 MeV/nucleon \citep{ace_data} taken together. 
(a) separately for each energy in $\sigma$'s, (b)
averaged for four energies in $\sigma$'s, and (c) relative.
\label{fig:equal_sigma}}
\end{minipage} \hfill
%
%
\begin{minipage}[tr]{\wcap}
\figcaption[f6.eps]{Deviation of propagated abundances (model DR II) from measured 
at 4.3, 5.6, 7.5, 10.6, 16.2 GeV/nucleon \citep[HEAO-3,][]{Engelmann90} 
given (a) separately for each energy in $\sigma$'s, (b)
averaged for all five energies in $\sigma$'s, and (c) relative.
\label{fig:HE_sigma} \medskip\vspace{2pt}}
\end{minipage}
\vspace{-0.5\baselineskip}
\end{figure*}

\placefigure{fig:equal_sigma}

Such effect as the atmospheric correction to the observed
antiproton flux is very important and may affect our results.
We discuss it in more details in Sect.~\ref{sec:conclusion}.

\section{APPLICATION TO NUCLEI UP TO Ni}

The DR I model gives an approximate fit to all elements (Fig.~\ref{fig:equal_sigma}).
The source elemental abundances are 
tuned (at a nominal 
reference energy of 100 GeV), by a least squares procedure, to
the abundances measured by HEAO-3 \citep{Engelmann90} at 7.5, 10.6,
16.2 GeV/nucleon combined with ACE 200 MeV/nucleon
data assuming modulation potential $\Phi=400$ MV.
At the chosen HEAO-3 energies the heliospheric modulation is weak 
(for the epoch 1980 we adopt $\Phi=800$ MV), and it is in the middle of
the logarithmic interval 0.6--35 GeV/nucleon covered by the HEAO-3
measurements; thus the systematic and statistical errors are
minimal. To the statistical errors of the ACE data, 
we added 5\% systematic error, which is a minimal conservative
estimate of the uncertainties in the measurements and modulation potential.

Fig.~\ref{fig:equal_sigma} shows the quality of the fit.
Fig.~\ref{fig:equal_sigma}a shows the deviation of the calculated abundances
from measurements at a given energy expressed in standard deviations.
Fig.~\ref{fig:equal_sigma}b shows the average deviation
$<\Sigma>=\frac1n\sum_{i=1}^n \frac{A^t_i-A^m_i}{\sigma_i}$,
where $A^t_i$, $A^m_i$ are the calculated and measured abundances
for the given energy, and $\sigma_i$ is the standard deviation.
Fig.~\ref{fig:equal_sigma}c shows the relative deviation of the calculated abundances
from measurements at a given energy.

The DR I fit is systematically 
low\footnote{This may be due to the errors in the production cross sections
employed in the calculations.}
at low energies (ACE) by as much as 15--30\% for elements $Z=11,13,19-25$.
It disagrees with high energy abundance of iron by $\sim20$\%, 
and by more than 50\% for \e{}{27}Co, and \e{}{28}Ni.
Meanwhile the high energy data taken separately are consistent within 5--10\%
(Fig.~\ref{fig:equal_sigma}c).
Because of this low energy discrepancy we consider further only 
the model DR II, where we allow Galactic CR and LB abundances to be different.
In this model, the low energy data are used to determine the LB source abundances.
DR II model provides the best fit to all data at the cost of extra free parameters.

\begin{figure*}[!thb] 
\psfig{file=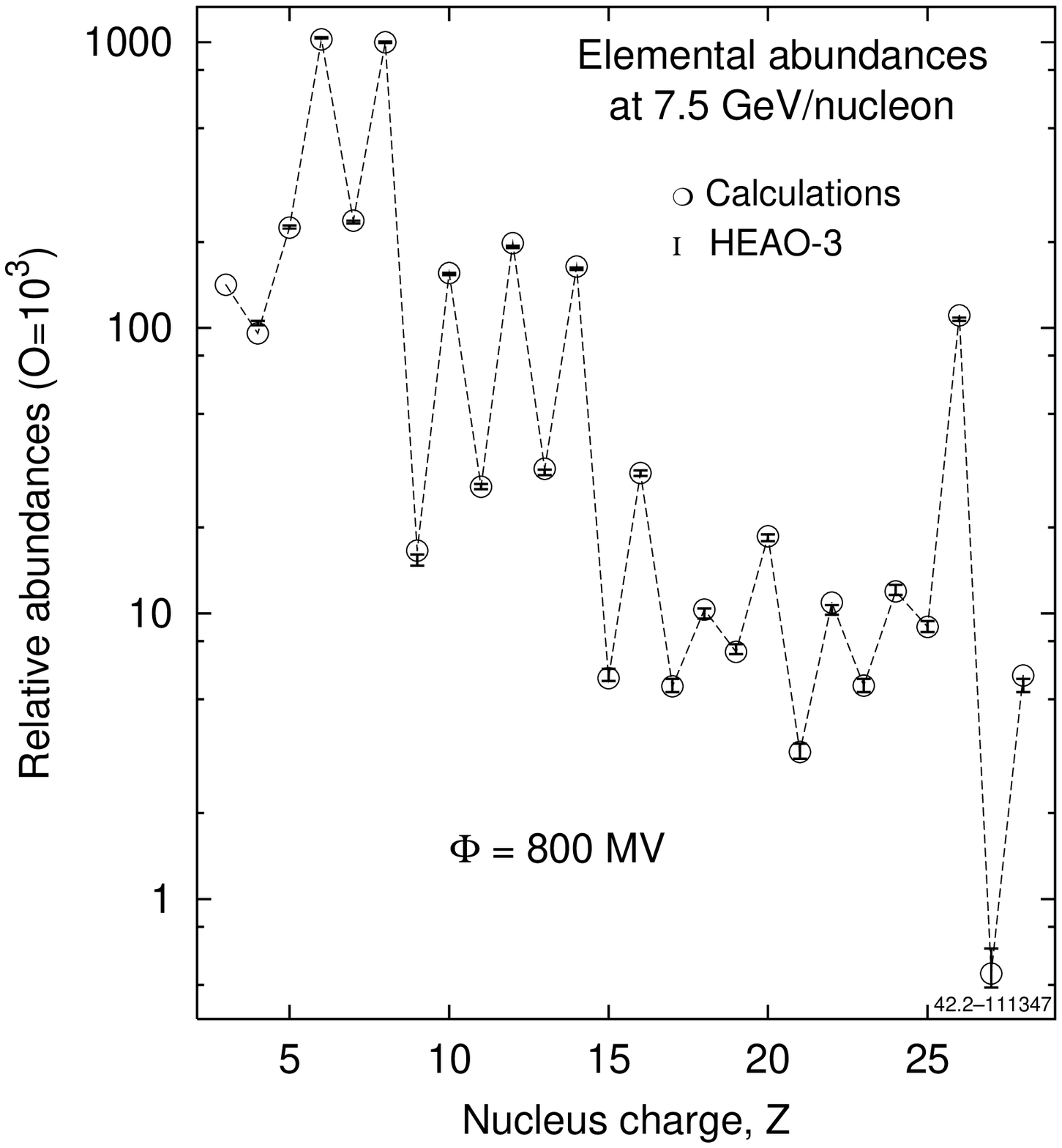,width=\twofigs,clip=}
\psfig{file=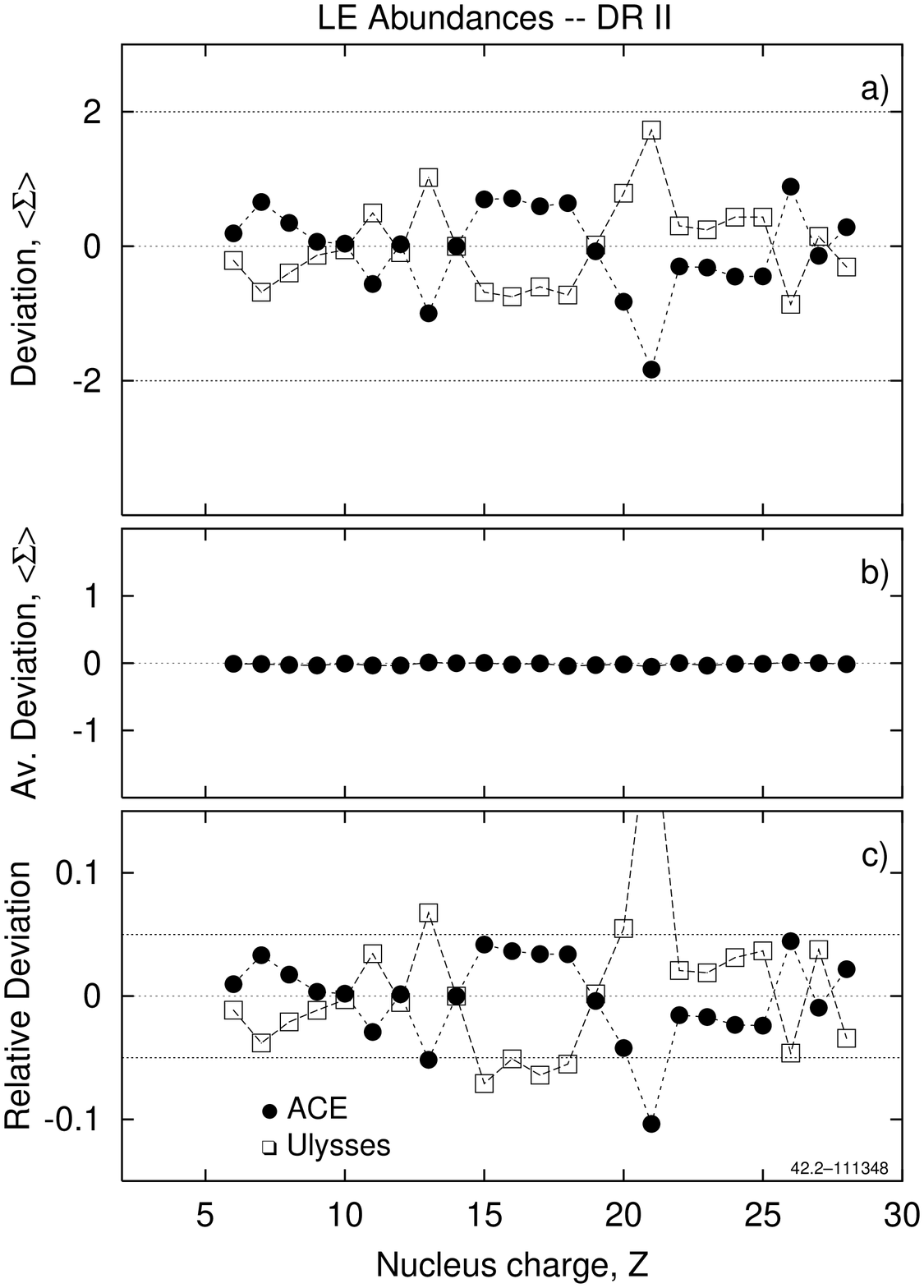,width=\twofigs,clip=}
\begin{minipage}[tl]{\wcap}
\figcaption[f7.eps]{Calculated propagated elemental abundances at 7.5 GeV/nucleon 
(model DR II).
Modulation - force field, $\Phi=800$ MV. Data: HEAO-3 \citep{Engelmann90}.
\label{fig:HE_Ab} \medskip\vspace{2pt}}
\end{minipage} \hfill
%
%
\begin{minipage}[tr]{\wcap}
\figcaption[f8.eps]{Deviation of propagated abundances (model DR II) from measured 
by ACE \citep{ace_data} and Ulysses \citep{ulysses} 
given (a) separately for each energy in $\sigma$'s, (b)
averaged for both sets of data in $\sigma$'s, and (c) relative.
\label{fig:LE_sigma}}
\end{minipage}
\vspace{-1\baselineskip}
\end{figure*}

\placefigure{fig:HE_Ab}

\enlargethispage*{2\baselineskip}

In the DR II model, the Galactic CR source elemental abundances are tuned (at a nominal 
reference energy of 100 GeV), by a least squares procedure, to
the abundances measured by HEAO-3 \citep{Engelmann90} at 5.6, 7.5, 10.6,
16.2 GeV/nucleon (Fig.~\ref{fig:HE_sigma}). Fig.~\ref{fig:HE_Ab}
shows calculated propagated abundances vs.\ HEAO-3 data at one 
particular energy 7.5 GeV/nucleon.
Relative isotopic abundances at the source are taken
equal to solar system abundances \citep{solar_isot_ab}.
The key point in the fitting procedure is to obtain the correct 
abundance of boron.

Fig.~\ref{fig:HE_sigma} shows the quality of the fit to high energy data,
where Figs.~\ref{fig:HE_sigma}a,b,c
show the deviation of the calculated abundances
from measurements at a particular energy expressed in standard deviations,
an average deviation, and the relative deviation of the calculated abundances
from measurements at a given energy, respectively.
The deviations from the data at any particular energy are almost 
all within $\sim5$\%. The calculated abundance of
\e{}{4}Be appears to be $\sim7$\% below the HEAO-3 data.
It is, however, the lightest nucleus measured by the apparatus, and its
measurements thus may be affected by systematic errors.
Some disagreement in calculated and measured abundances of \e{}{9}F and \e{}{23}V 
as seen in Fig.~\ref{fig:HE_sigma}b
is caused by overproduction at only one energy point, 
7.5 GeV/nucleon in case of \e{}{9}F and 
10.6 GeV/nucleon in case of \e{}{23}V, while at
other energies calculations agree well with data.
10\% overproduction of \e{}{24}Cr is seen only at
one energy 10.6 GeV/nucleon. Compared to the calculations,
measurements of \e{}{19}K and \e{}{27}Co are particularly
scattered; \e{}{27}Co is the least abundant element for $Z<29$ and
its abundance in CR is measured with large error bars. \e{}{28}Ni
is at the end of the nucleus charge interval measured by HEAO-3
and probably its measurement is also affected by the systematic errors.
In general, the deviations in measurements are larger for the least 
abundant nuclei, which is not surprising. 

The LB elemental
abundances are tuned simultaneously with spectra using the
low energy part of the B/C ratio and isotopic abundances at 200 MeV/nucleon 
from ACE \citep{ace_data} and at 185 MeV/nucleon from Ulysses \citep{ulysses}.
For many elements ACE and Ulysses abundances differ by 10\% (Fig.~\ref{fig:LE_sigma}c).
For this reason, to the statistical
errors shown we added 5\% systematic error. 
In the same way as for the high energy data, Fig.~\ref{fig:LE_sigma} 
shows the quality of the fit to low energy data by 
ACE and Ulysses.

\placetable{table3}


\begin{table*}[!bh]
\vspace{-2\baselineskip}
\tablecolumns{4}
\tablewidth{-10mm}
\tablecaption{Elemental abundances\tablenotemark{a}.
\label{table3}}

\begin{tabular}{lccc lccc}

\tablehead{
\colhead{Z} & 
\colhead{Solar System} & 
\colhead{LB Sources} & 
\colhead{Galactic Sources}&
\colhead{Z} & 
\colhead{Solar System} & 
\colhead{LB Sources} & 
\colhead{Galactic Sources}
}
\startdata
   6 &$   9.324    $ &$   3.850    $ &$   4.081    $ &
  18 &$   7.070\times10^{-2}$ &$   3.283\times10^{-2}$ &$   1.915\times10^{-2}$\\ 
   7 &$   2.344    $ &$   6.500\times10^{-1}$ &$   3.022\times10^{-1}$ &
  19 &$   3.718\times10^{-3}$ &$   2.317\times10^{-2}$\tablenotemark{*} &$   6.392\times10^{-3}$\\ 
   8 &$   19.04    $ &$   6.317    $ &$   5.235    $ &
  20 &$   6.451\times10^{-2}$ &$   8.333\times10^{-2}$ &$   5.887\times10^{-2}$\\ 
   9 &$   8.901\times10^{-4}$ &$          0.$ &$          0.$ &
  21 &$   4.169\times10^{-5}$ &$   1.367\times10^{-2}$\tablenotemark{*} &$   1.730\times10^{-4}$\\ 
  10 &$   3.380    $ &$   6.667\times10^{-1}$ &$   6.328\times10^{-1}$ &
  22 &$   2.958\times10^{-3}$ &$   5.817\times10^{-2}$\tablenotemark{*} &$   3.166\times10^{-3}$\\ 
  11 &$   6.028\times10^{-2}$ &$   9.767\times10^{-2}$ &$   3.575\times10^{-2}$ & 
  23 &$   2.817\times10^{-4}$ &$   2.433\times10^{-2}$\tablenotemark{*} &$          0.$\\ 
  12 &$   1.070    $ &$   1.385    $ &$   1.050    $ &
  24 &$   1.318\times10^{-2}$ &$   6.200\times10^{-2}$ &$   2.481\times10^{-2}$\\ 
  13 &$   8.310\times10^{-2}$ &$   1.583\times10^{-1}$ &$   7.794\times10^{-2}$ & 
  25 &$   6.901\times10^{-3}$ &$   1.800\times10^{-2}$ &$   2.309\times10^{-2}$\\ 
  14 &$   1.       $ &$   1.       $ &$   1.       $ &
  26 &$   8.901\times10^{-1}$ &$   7.767\times10^{-1}$ &$   9.661\times10^{-1}$\\ 
  15 &$   7.944\times10^{-3}$ &$   5.667\times10^{-3}$ &$   1.041\times10^{-2}$ & 
  27 &$   2.344\times10^{-3}$ &$   4.500\times10^{-3}$ &$   1.773\times10^{-3}$\\ 
  16 &$   6.028\times10^{-1}$ &$   1.000\times10^{-1}$ &$   1.425\times10^{-1}$ & 
  28 &$   5.014\times10^{-2}$ &$   3.567\times10^{-2}$ &$   5.591\times10^{-2}$\\ 
  17 &$   8.901\times10^{-3}$ &$   4.667\times10^{-3}$ &$   4.047\times10^{-3}$\\ 
\enddata
\footnotesize
\tablenotetext{a}{Normalized to Si=1.\vspace{-0.3\baselineskip}} 
\tablenotetext{*}{Upper limit.}
\vspace{-1\baselineskip}
\end{table*}

\begin{figure*}[!thb] 
\epsscale{\fsizetwo}
\plottwo{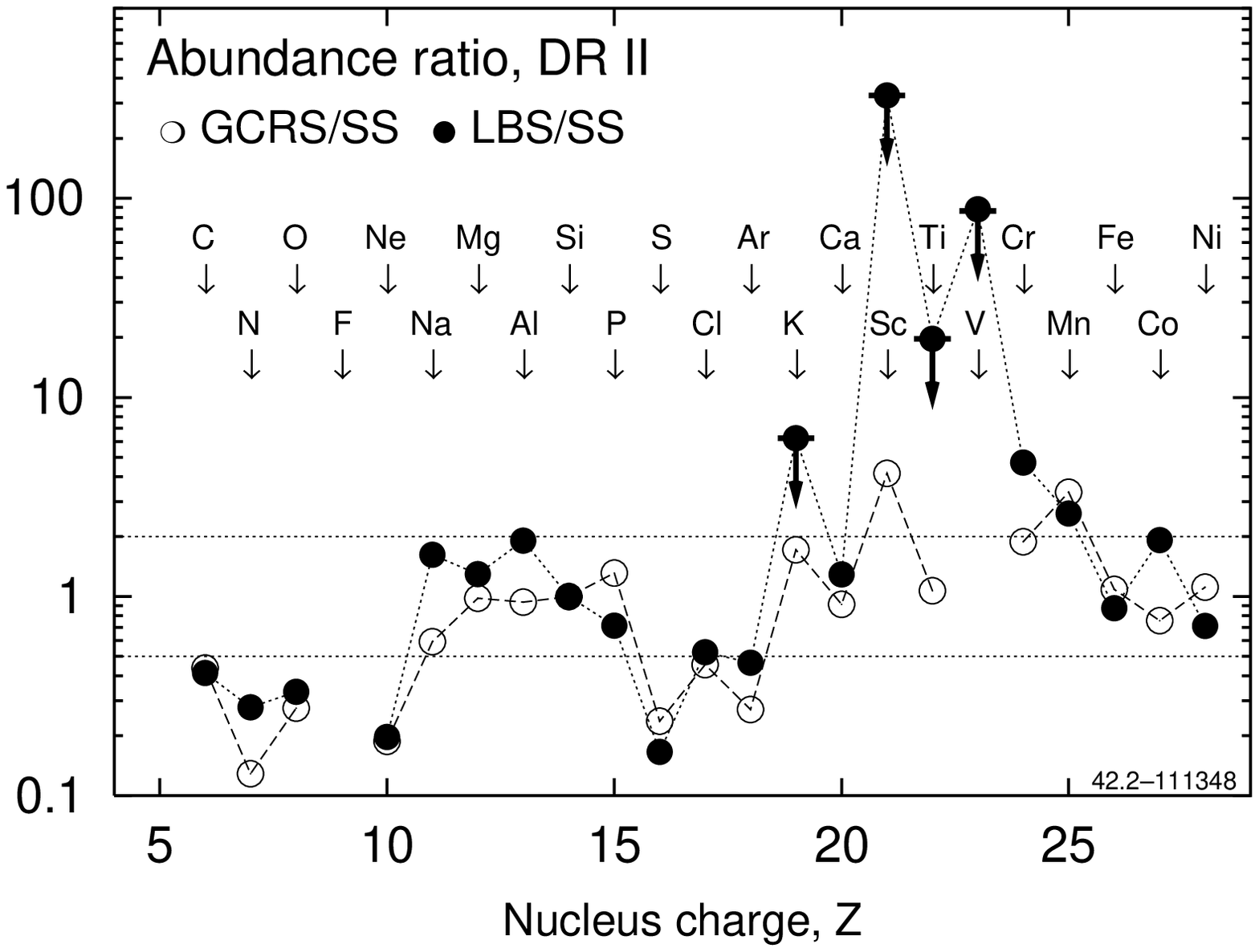}{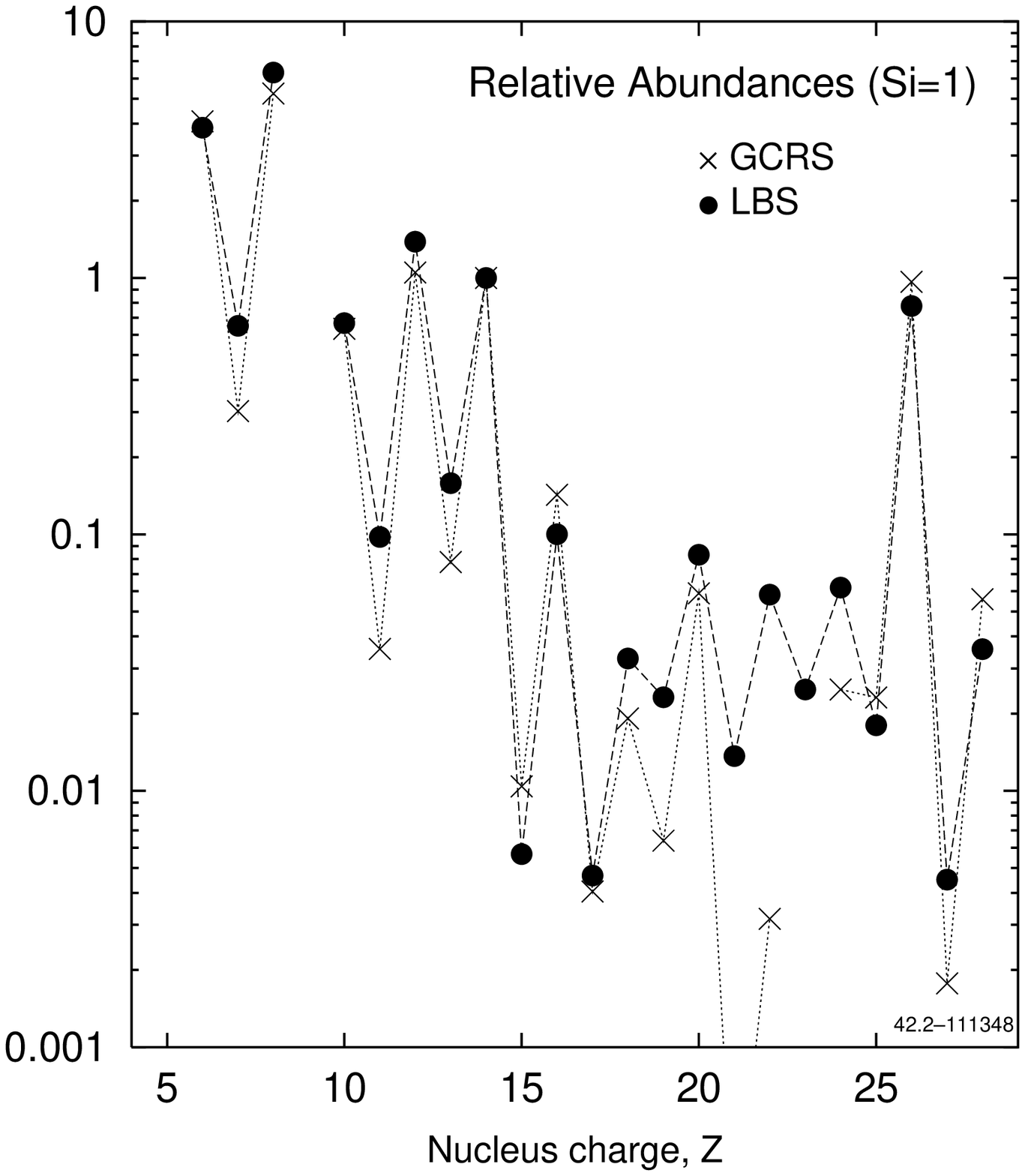}
\caption[f9a.eps,f9b.eps]{\emph{Left:} Derived abundance ratios (model DR II)
Galactic-CR-source/Solar-System (GCRS/SS) 
and LB-source/Solar-System (LBS/SS), normalized to silicon. Relative
abundances for K, Sc, Ti, V are shown as upper limits.
Solar system abundances from
\citet{solar_elem_ab}. The dotted lines plotted at 1/2 and 2.
\emph{Right:} Derived Galactic CR source and LB source abundances normalized to silicon.
\label{fig:Ab-ratio}}
\end{figure*}

\placefigure{fig:Ab-ratio}

The fitting procedure is also influenced by the adopted value of the 
modulation potential.
We found that applying the 
following pairs of modulation potentials yields
almost the same LB elemental abundances: 
400 MV for ACE and 700 MV for Ulysses, 
450 MV -- ACE and 600 MV -- Ulysses, and 500 MV -- ACE and 500 MV -- Ulysses.
Other combinations of modulation potentials make the 
quality of the fit worse.
The data are shown to deviate from calculations in
both directions, which mean that we are unlikely to
introduce essential systematic error by assuming a wrong value of the
modulation potential.

\smallskip
\section{ABUNDANCES IN COSMIC RAYS AND COSMIC RAY SOURCES} \label{sec:results}
\subsection{Elemental abundances in cosmic ray sources}

\begin{figure*}[!thb] 
\plottwo{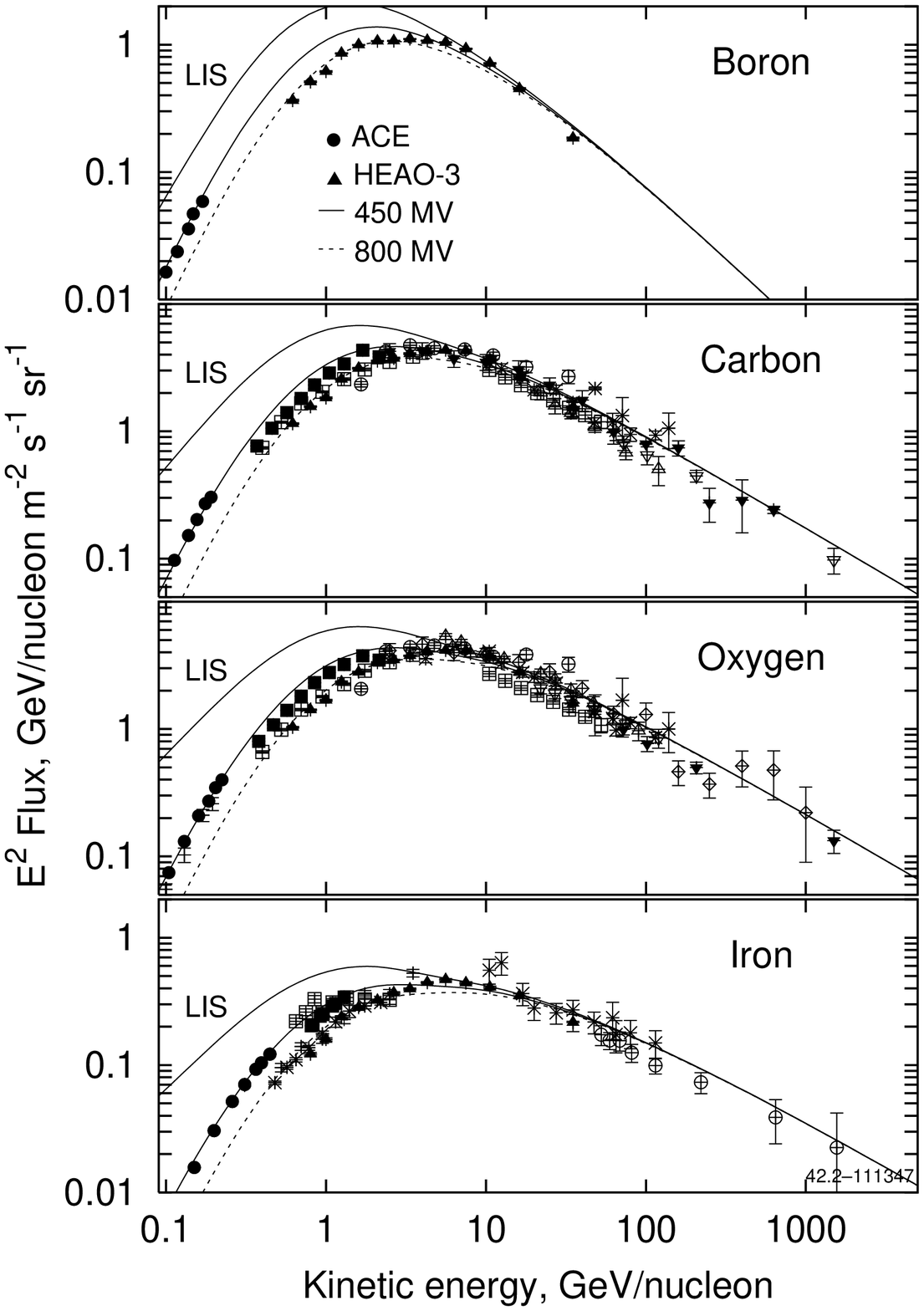}{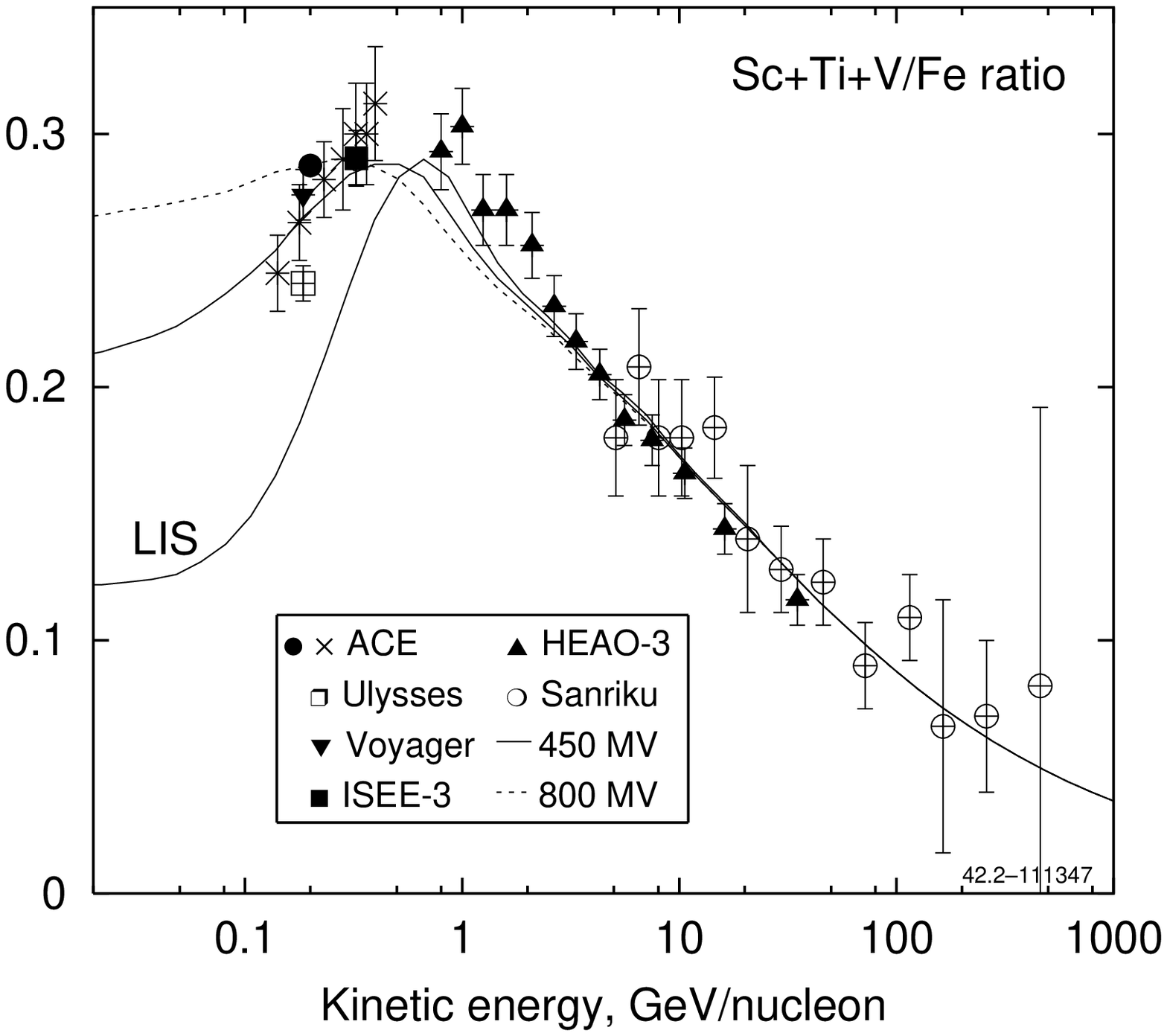}
\begin{minipage}[tl]{\wcap}
\figcaption[f10.eps]{Spectra of boron, carbon, oxygen, and iron 
(from top to bottom) calculated with LB contribution (model DR II).
Upper curves - LIS, lower curves -- modulated using force field
approximation ($\Phi=450$ MV -- solid curves, $\Phi=800$ MV -- dashes). 
Data: ACE \citep{davis,davis01},
HEAO-3 \citep{Engelmann90}, for other references see
\citet{StephensStreitmatter98} (symbols are changed). 
\label{fig:B-Fe}}
\end{minipage} \hfill
%
%
\begin{minipage}[tr]{\wcap}
\figcaption[f11.eps]{
Sub-Fe/Fe ratio calculated with LB contribution (model DR II),
and with $E_b=500$ MeV/nucleon.
Lower curve -- LIS, upper curves -- modulated (force field, $\Phi=450$ and 800 MV). 
Data: ACE \citep{ace_data,davis}, Ulysses \citep{ulysses}, Voyager
\citep*{L97}, ISEE-3 \citep{leske}, HEAO-3 \citep{Engelmann90}, Sanriku \citep{Sanriku}.
\label{fig:sub-Fe}}
\end{minipage}
\vspace{-1\baselineskip}
\end{figure*}

The DR II model with an LB component shows good overall 
agreement with data including secondary to primary ratios, spectra, 
and abundances. 
The derived Galactic CR source abundances and LB source abundances are
given in Table \ref{table3} and
plotted in Fig.~\ref{fig:Ab-ratio} relative to the solar system abundances.  
Spectra of boron, carbon, oxygen, and iron are shown in Fig.~\ref{fig:B-Fe}
for two modulation levels, 450 and 800 MV.
In case of carbon, the normalization coefficient
in the LB component (eq.\ [\ref{eq1}]) is fixed as $a(6,12)=6.35\times10^{-4}$
cm$^{-2}$ s$^{-1}$ sr$^{-1}$ for $\eta=1$. The calculated sub-Fe/Fe ratio is
plotted in Fig.~\ref{fig:sub-Fe}. Since the elemental abundances are
tuned at both high and low energies, it agrees well with data.
In the intermediate region at $\sim1$ GeV/nucleon, the agreement
could be improved by taking $E_b$ smaller than currently assumed
500 MeV 
(e.g., adopting $E_b=400$ MeV may raise it by $\sim10$\%, 
similar to the change
in the B/C ratio, Fig.~\ref{fig:BC300} right).


The important result (Fig.~\ref{fig:Ab-ratio}) is that CR source
and LB source abundances of all major elements 
(\e{}{6}C, \e{}{8}O, \e{}{10}Ne, \e{}{12}Mg, \e{}{14}Si, \e{}{16}S, \e{}{18}Ar, 
\e{}{20}Ca, \e{}{26}Fe, \e{}{28}Ni) are in good agreement with each other.
Abundances of the Si group, \e{}{20}Ca, \e{}{26}Fe, \e{}{28}Ni are near the 
Solar System abundances.
Abundances of other elements in Galactic CR and LB sources 
are mostly consistent with each other, and with
solar system abundances, within a factor of 2.
Relative to silicon, \e{}{6}C, \e{}{7}N, \e{}{8}O, \e{}{10}Ne, \e{}{16}S 
are underabundant both in CR source and LB. 
This corresponds to the well-known first-ionization-potential or
volatily correlation \citep[see, e.g.,][]{meyer}. 
Nitrogen in CR sources and LB sources differs by a factor
of $\sim3$, which may be connected with production cross section
errors affecting propagation of Galactic CR component (see discussion in 
Appendix~\ref{cs}). 

\placefigure{fig:B-Fe}

Secondary nuclei \e{}{19}K, \e{}{21}Sc, \e{}{22}Ti, \e{}{23}V 
appear to be overabundant in the LB sources relative to the solar system
(shown as upper limits in Fig.~\ref{fig:Ab-ratio}a), though the derived
absolute LB abundances are not large. The derived LB abundance  
of \e{}{22}Ti does not exceed that of \e{}{24}Cr while the derived
abundances of \e{}{21}Sc and \e{}{23}V are not larger than that of \e{}{25}Mn
(Fig.~\ref{fig:Ab-ratio}b).
One possible reason for this excess is the uncertainty in
the production cross sections, which is especially large for these nuclei. 
Sometimes there is no measurement at all;
in this case one can use only phenomenological
systematics, which are frequently wrong by a factor of two or even more,
and/or predictions by Monte Carlo codes.
Often, there is only one measurement at $\sim600$ MeV/nucleon,
which has to be extrapolated in both 
directions\footnote{Only the following reactions are 
well measured \citep[see compilation by][]{t16lib}, on a sample of natural 
iron consisting mostly of \e{56}{}Fe: 
$p+$natFe $\to$ \e{46,47}{\ \ \ 21}Sc, \e{48}{23}V, \e{48,51}{\ \ \ 24}Cr; 
we use our fits to these data. Reactions producing other isotopes of 
\e{}{21}Sc, \e{}{22}Ti, \e{}{23}V
by \e{56}{26}Fe, \e{55}{25}Mn, and \e{52}{24}Cr have only one or two measurements.
There is no data available on the production of \e{}{21}Sc, \e{}{22}Ti, \e{}{23}V by 
\e{54}{26}Fe, \e{53}{25}Mn, and
\e{}{24}Cr isotopes (except \e{52}{24}Cr), on production of \e{}{21}Sc, \e{}{22}Ti, 
by \e{49}{23}V, and on production of \e{}{21}Sc by \e{}{22}Ti. 
These poorly known cross sections  
contribute to errors on the production of sub-Fe elements at low energies.}.
This allows only a nearly flat Webber-type or Silberberg-Tsao-type
extrapolations while the real cross sections
usually have large resonances below several hundred MeV, and 
decrease with energy above a few GeV \citep[see, e.g.,][]{MMS}.
We note that \citet{davis} used semiempirical cross sections based on
\citeauthor{W-code} and also predicted fluxes of sub-Fe elements which are too low.  

An estimate of the overall error, which is reflected in the derived LB source
abundances, can be obtained by assuming the complete 
absence of 
\e{}{19}K, \e{}{21}Sc, \e{}{22}Ti, \e{}{23}V 
in the LB source (shown by crosses in Fig.~\ref{fig:Ab-ratio}).
In this case
the discrepancy between the calculated propagated CR abundances 
of \e{}{19}K, \e{}{21}Sc, \e{}{22}Ti, \e{}{23}V, and those measured
is below 20\%, and
can be \emph{removed} by allowing the production cross 
sections to increase at low energies by $\sim15-20$\%, which seems plausible.

Another possibility is errors in flux measurements of the rare CR species.
Fig.~\ref{fig:LE_Ab} shows the calculated 
abundances\footnote{Calculated Li abundance in the plot shows only secondary 
lithium produced in CR.} 
tuned at low energies to the ACE and Ulysses
data. 
Ulysses and ACE measurements are not always in agreement.
Note that even for such an abundant nucleus as iron, which is the
main contributor to the sub-Fe group, the discrepancy 
exceeds 10\%, while the disagreement in abundance of \e{}{21}Sc is 
$\sim30$\%.

The derived source overabundance of sub-Fe elements in the LB could 
also in principle arise from composition differences between the ISM in the LB and
solar or Galactic average ISM. This is suggested by 
the fact that the relative abundances of secondary elements in the LB
sources are systematically larger than in the Galactic
CR sources (Fig.~\ref{fig:Ab-ratio}).
However, the factors required in case of, e.g., \e{}{22}Ti
(Ti/Fe $\sim5$\% compared to solar or SN 0.1\%)
appear much larger than could reasonably
be expected even for unusual SN types. 
A similar effect may be caused by a specific
correlation between source and gas density distributions \citep{soutoul}.

\subsection{Isotopic distributions in cosmic rays}\label{sec:isotopes}
Be and B isotopes are assumed all secondary, thus there is no
possibility to tune them. The DR II model calculation shows perfect agreement
with the data on relative isotopic abundances of Be and B
(Fig.~\ref{fig:BeBC}). This is in contrast with a ``standard''
reacceleration model, where we obtained a 15\% discrepancy with relative
abundances of \e{7}{}Be and \e{9}{}Be isotopes \citep{SM01}.

Abundances of stable isotopes of other elements are not very conclusive
because they are present in the sources, but O and Si isotopic distributions
still agree very well with data \citep*{w96,w97,d96,h96,ace_data}
assuming only \e{16}{}O and \e{28}{}Si isotopes are present in the LB component.
C and N isotopic distributions
do not agree too well (Fig.~\ref{fig:BeBC}), but this may point to a problem with
cross sections.
The calculated ratio \e{13}{}C/\e{12}{}C $\sim0.11$ at 120 MeV/nucleon
($\Phi= 500$ MV) in the model with LB contribution
is still a factor $\sim1.5$ too large compared to the measured value, $0.0629\pm0.0033$
\citep[Voyager 50--130 MeV/nucleon,][]{w96} and $0.078\pm0.011$ 
\citep[Ulysses 100--200 MeV/nucleon,][]{d96},
which may be connected in part with overproduction
of \e{13}{}C by \e{15}{}N. 
(A discussion of the cross-section uncertainties for C and N isotopes
is given in Appendix \ref{cs}.)
If we replace the cumulative cross section \e{15}{}N$+p\to$\e{13}{}C
with cross section \e{14}{}N$+p\to$\e{13}{}C, the calculated ratio 
\e{13}{}C/\e{12}{}C will be lowered by 10\% as estimated (see Appendix \ref{cs}
and Fig.~\ref{fig:cn}). Assuming the absence of the isotope \e{13}{}C
in the Galactic CR sources gives another 10\% reduction. Altogether
these corrections yield \e{13}{}C/\e{12}{}C $\sim0.09$, close to the data.

\begin{figure*}[!thb]
\epsscale{\fsizetwo}
\plottwo{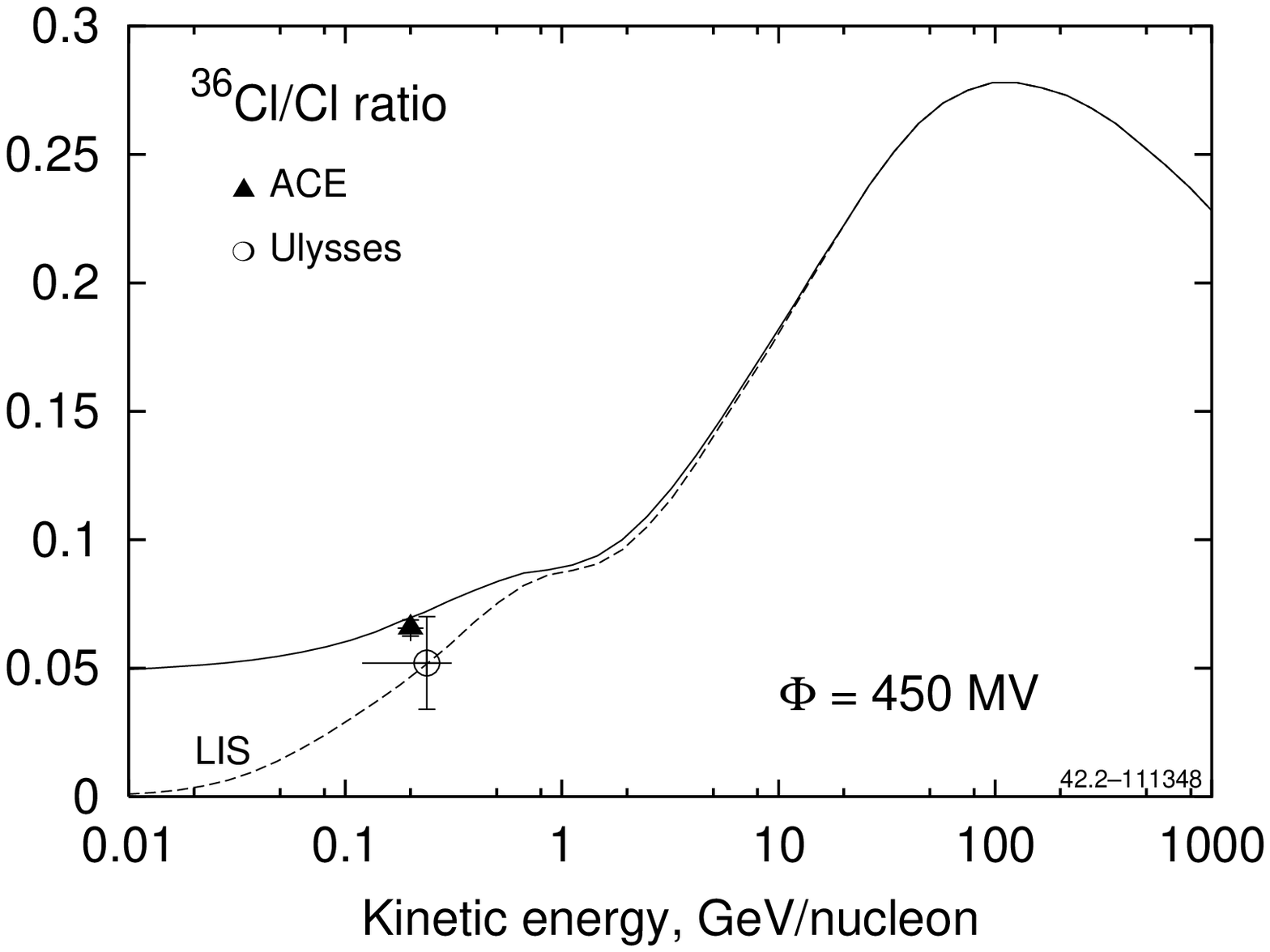}{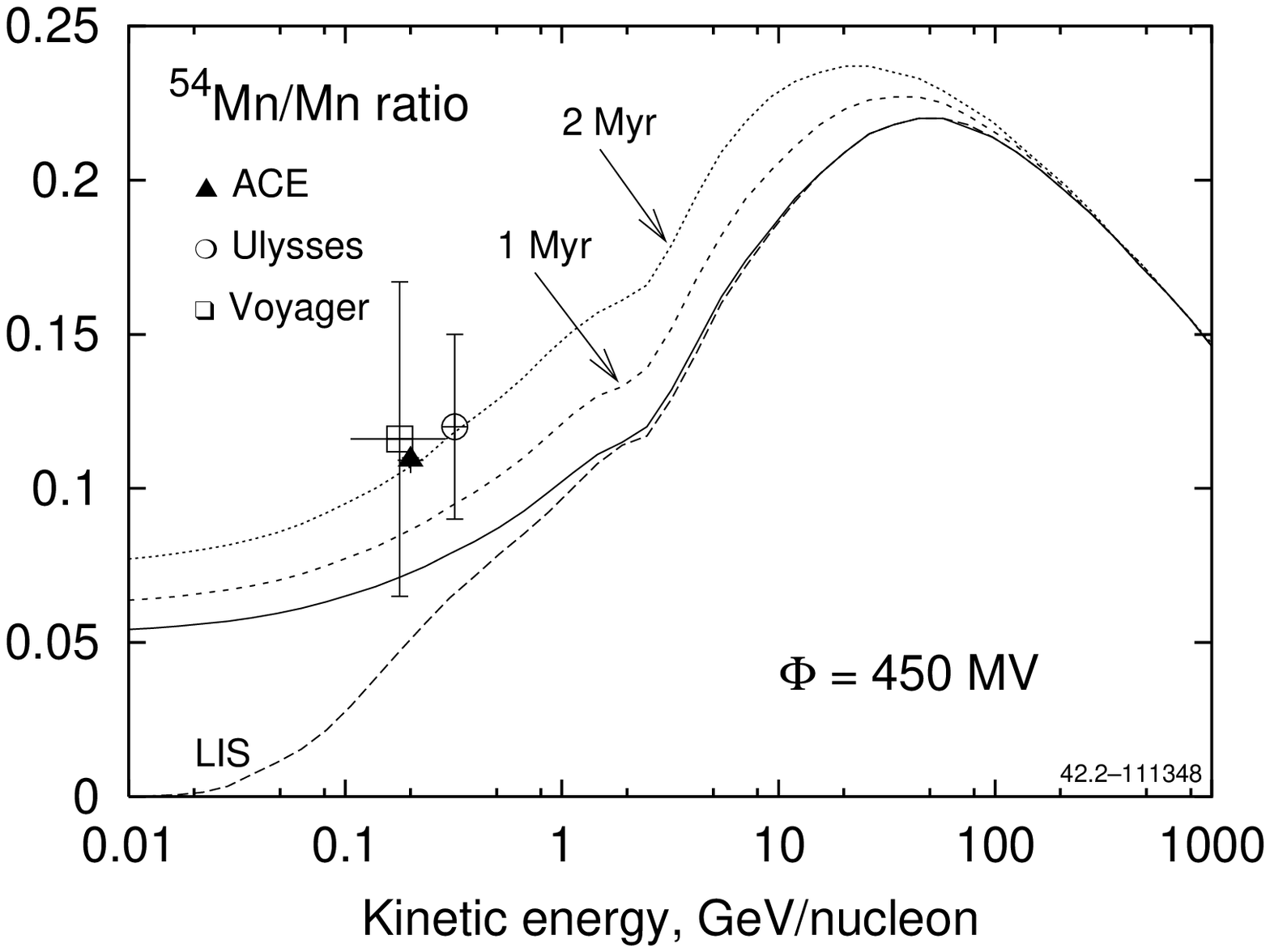}
\caption[f14a.eps,f14b.eps,f14c.eps,f14d.eps]
{``Radioactive clocks'' isotopic ratios in CR, 
as calculated in DR II model for $z_h=4$ kpc.  
Dashed curve -- interstellar (LIS), solid curve -- modulated ($\Phi=450$ MV).
Mn plot shows also the ratio calculated for two half-lives, 
$t_{1/2}=1$ Myr and 2 Myr,
that are different from the adopted $t_{1/2}=0.63$ Myr.
Be data: Ulysses \citep{C98}, Voyager \citep{voyager}, ACE \citep{B99}, ISOMAX \citep{H01,N01};
Al data: Ulysses \citep{Al26_ulysses}, ACE \citep{ace_data};
Cl data: Ulysses \citep{Cl36_ulysses}, ACE \citep{ace_data};
Mn data: Ulysses \citep{Mn54_ulysses}, Voyager \citep{L97}, ACE \citep{ace_data}.
\label{fig:Be}}
\end{figure*}

Fig.~\ref{fig:Be} shows calculated \e{10}{}Be/\e{9}{}Be,
\e{26}{}Al/\e{27}{}Al, \e{36}{}Cl/Cl, \e{54}{}Mn/Mn ratios,
usually used as ``radioactive clocks'' in CR, for a halo
size $z_h=4$ kpc vs.\ data. In case of Be, Al, Cl, 
the agreement with the most accurate low energy data by ACE
is very good and all the ratios are consistent with 
%
\bigskip
\centerline{\psfig{file=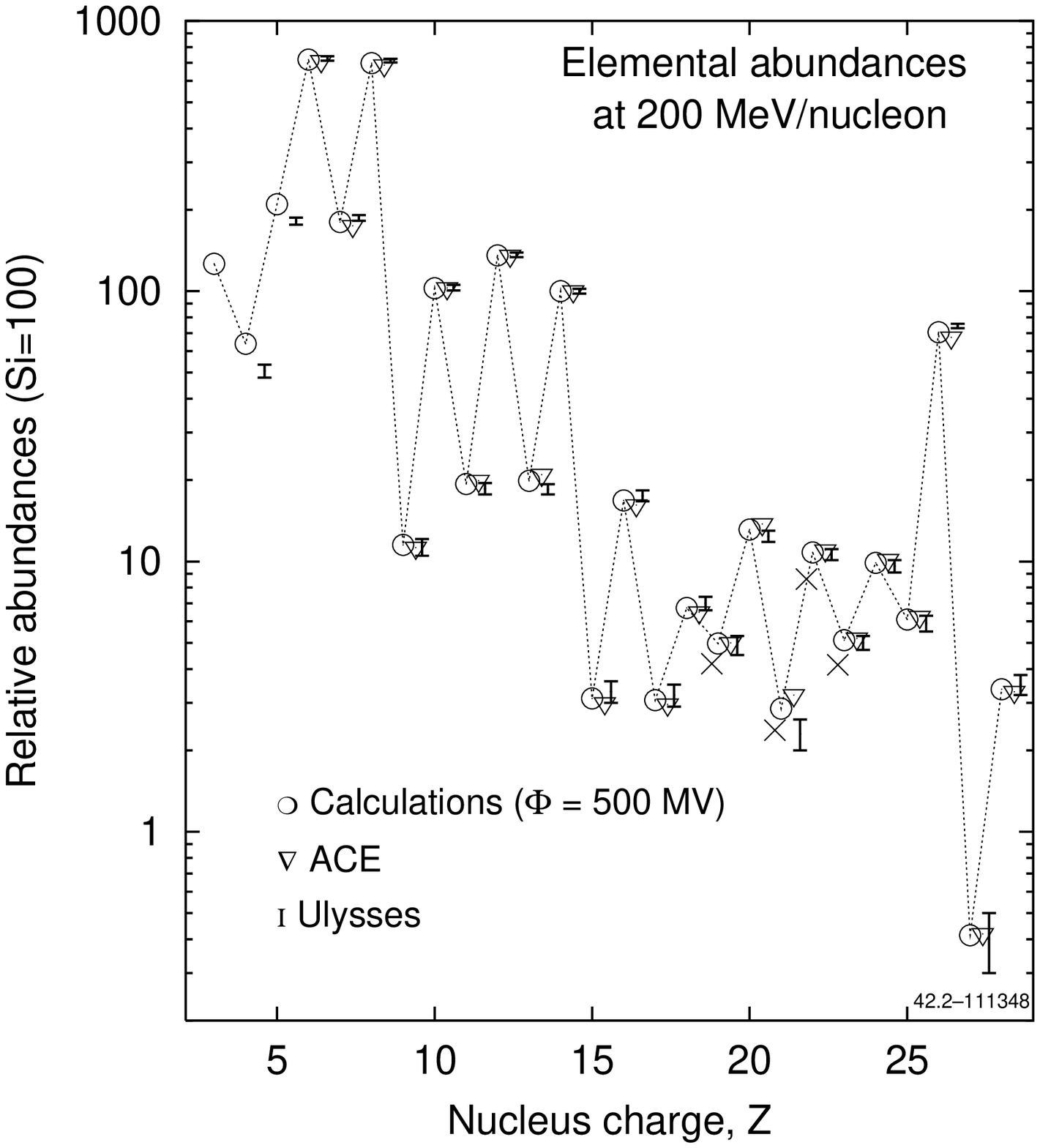,width=\twofigs,clip=}}
\figcaption[f12.eps]{Propagated elemental abundances at 200 MeV/nucleon with
LB contribution (model DR II). Crosses show the calculated abundances 
assuming no K, Sc, Ti, V in the LB source.  
Data: ACE \citep{ace_data}, Ulysses \citep{ulysses}. \bigskip\bigskip
\label{fig:LE_Ab} }
\noindent
each other
indicating that $z_h=4$ kpc is a good estimate. 
Higher energy data by ISOMAX (Be) are also consistent with 
calculations considering the large error bars.

The \e{54}{}Mn/Mn ratio indicates a somewhat smaller halo, but this may be related
to uncertainty in its half-life and/or production cross section. 
The half-life of \e{54}{}Mn
against $\beta^-$ decay is the most uncertain among the four
radioactive isotopes -- it is only one which is not measured
directly. It is derived indirectly based on $\beta^+$
decay branch half-life, which yields an estimate $t_{1/2}(\beta^-)=
6.3\pm1.3[{\rm stat}]\pm1.1[{\rm theor}]\times10^5$ yr \citep{Mn_lifetime}.
In case only the half-life is wrong,
to get the \e{54}{}Mn/Mn ratio consistent with lighter element ratios
and with ACE data (for $z_h=4$ kpc) requires 
$t_{1/2}(\beta^-)\sim2$ Myr (Fig.~\ref{fig:Be}). 
Another estimate of \e{54}{}Mn partial half-life
based on CR propagation calculations gives $\sim1-2$ Myr 
\citep{Mn54_ulysses}.
Apart from the half-life, a possible source of errors
can be production cross sections of Mn isotopes.
The fact that the propagated isotopic abundance of \e{53}{}Mn,
\e{53}{}Mn/Mn$=0.50$, is correct
(\e{53}{}Mn is a K-capture isotope that is absent in the ISM)
indicates that some important production channels of \e{54,55}{}Mn
may be not calculated correctly. (For instance, 
in case of stable \e{55}{}Mn,
we have a freedom to choose its abundance in the LB component
at low energies, which may compensate
for underproduction of this isotope in CR.
Meanwhile, this LB \e{55}{}Mn does not produce any \e{54}{}Mn.)
Only the reaction \e{nat}{}Fe$+ p\to$\e{54}{}Mn on a natural sample
of Fe has been measured well enough \citep[see compilation by][]{t16lib}. 
The cross section 
%
\centerline{\psfig{file=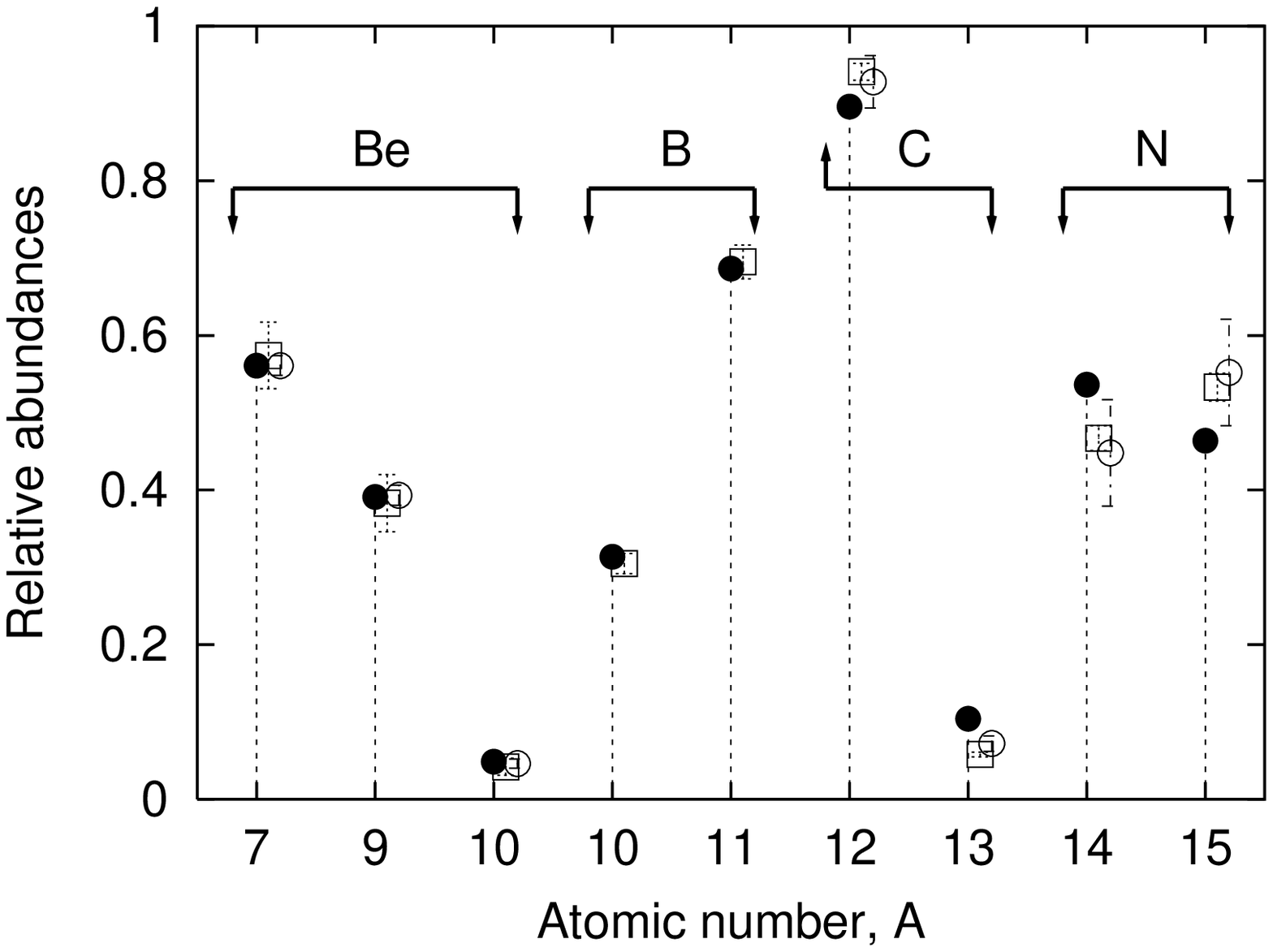,width=\twofigs,clip=}}
\figcaption[f13.eps]{Be, B, C, and N isotope distribution as calculated
in DR II model (solid circles) at $\sim70-150$ MeV/nucleon and a modulation potential 
$\Phi=450-500$ MV compare to the data.
Data: Be: Ulysses \citep{C98}, Voyager \citep{voyager}, B: Voyager \citep{voyager},
C,N: Voyager \citep{w96}, Ulysses \citep{d96}. \bigskip
\label{fig:BeBC} }
\noindent
\e{55}{}Mn$+ p\to$\e{54}{}Mn has only one point measured by
\citet{We98prc}, besides, it seems too low compared to similar neutron
knock out reactions on isotopes of Fe and Cr.
Reactions \e{55}{}Fe$+ p\to$\e{54}{}Mn, \e{57}{}Fe$+ p\to$\e{54}{}Mn
are not measured at all. Similarly, only one \e{55}{}Mn major production 
cross section \e{56}{}Fe$+ p\to$\e{55}{}Mn has even one point measured.

\placefigure{fig:Be}

\begin{figure*}[!thb] 
\epsscale{\fsizetwo}
\plottwo{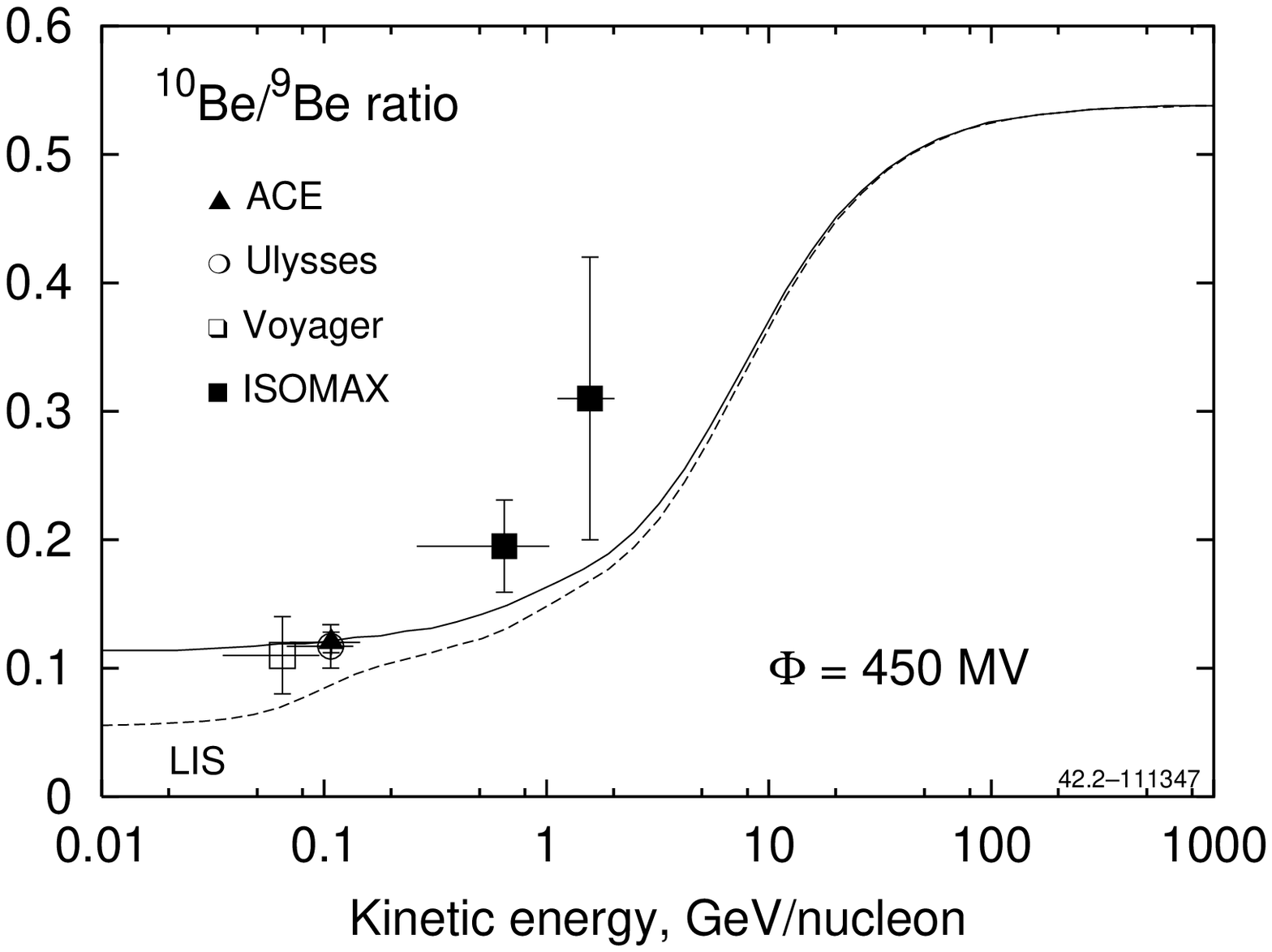}{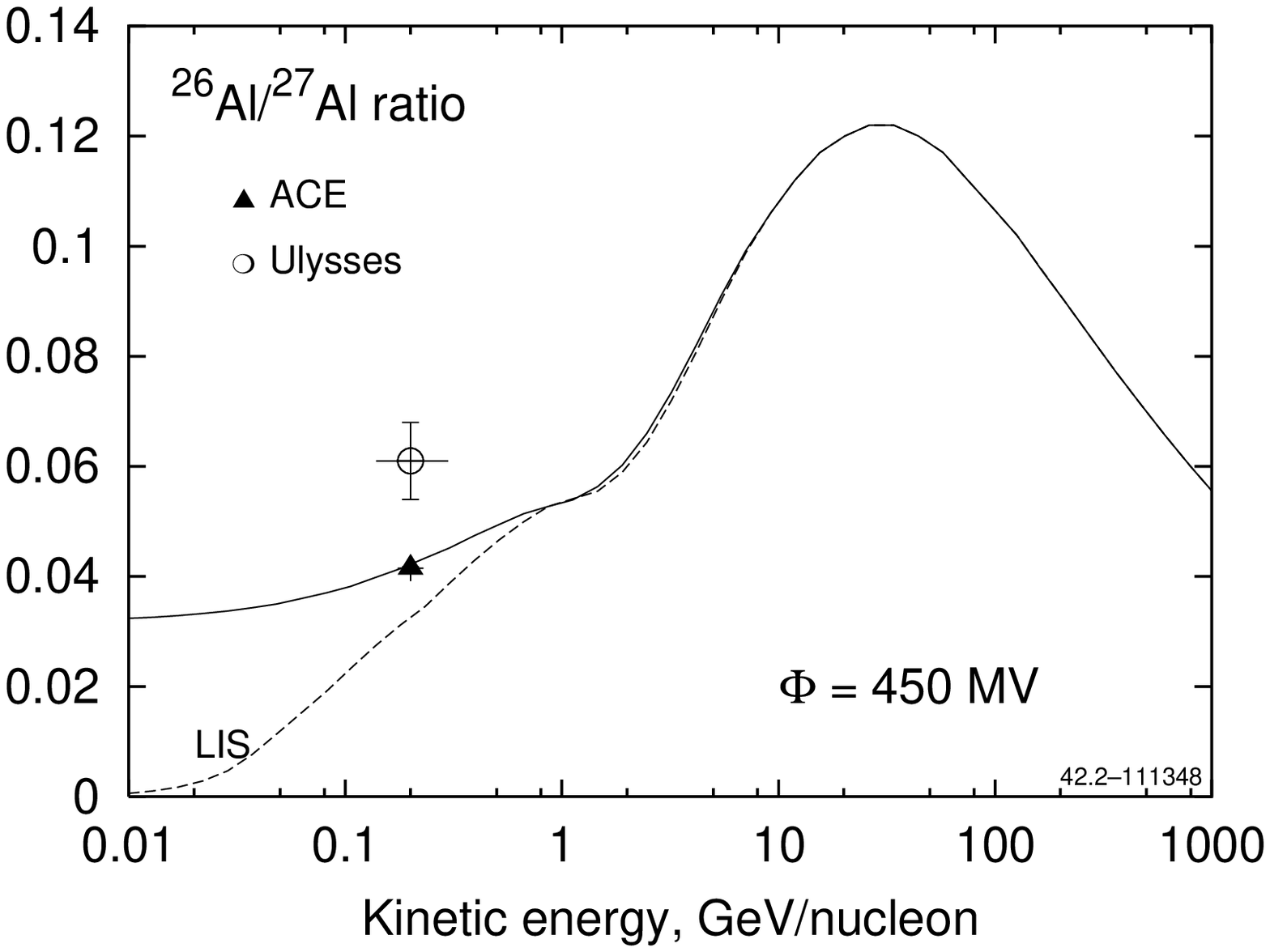}
\end{figure*}

\section{DISCUSSION}\label{sec:discussion}
The proton spectrum at low energies still remains uncertain.
The only secondaries produced below the antiproton production threshold
are positrons and \grays. However, the positron spectrum \emph{alone} 
cannot provide conclusive information on the proton
spectrum on a large scale because (i)
the large energy losses of positrons mean that the positron spectrum by its 
nature is local (ii) and there are possibly sources of primary positrons such
as pulsars. Diffuse \grays\ can provide a tool to test the
spectrum of protons in distant regions, but the \emph{a priori}
unknown contribution of electrons via inverse Compton and
bremsstrahlung complicates the picture.
A test of the He spectrum at energies below $\sim10$ GeV/nucleon 
can be made using the CR deuteron and \e{3}{}He measurements similar
to what was done in this paper for heavier nuclei.
We plan to address this issue in future work.

We should mention that there is 
another possibility to get the correct antiproton flux in reacceleration
models, which is to introduce an additional proton component at
energies up to approximately 20 GeV. The latter energy is above the
antiproton production threshold and effectively produces antiprotons
at $\sim 2$ GeV and below.  The intensity and spectral shape of this
component could be derived by combining restrictions from antiprotons
and diffuse \grays. 
Interestingly this kind of spectrum was used in our HEMN model 
\citep*[hard electrons and modified nucleons,][]{SMR00} to match the
spectrum of diffuse \grays\ as observed by EGRET \citep{hunter97}. 
The advantage of this approach
is that the diffuse \grays\ which we observe carry 
information on the large-scale Galactic spectrum of CR (producing antiprotons)
while particles we measure may reflect only the local region.

One more
(non-standard) interpretation is that the solar modulation is weaker
than assumed, and this would eliminate the need for a LB
component. 
(A cornerstone of the current theories of heliospheric modulation is 
the local interstellar spectrum, which is not known but taken \emph{a priori}.)
With a modulation potential as small as $\sim200$ MV one can
obtain a consistent reacceleration model combining B/C, antiprotons,
and other species simultaneously. To get an agreement with 
nucleon spectral data, the injection spectra in such a model
should be flatter at low energies than the usually adopted power-law in 
rigidity.

Recently there has appeared some indication that the atmospheric contribution
to the antiproton flux measured in the upper atmosphere is underestimated.
Monte Carlo simulations of the hadron cascade development in the upper
atmosphere have shown that the antiproton flux induced by $pA$-reactions
on air nuclei is larger, \emph{at least}, by $\sim$30\%
\citep*{atmospheric_pbars} compared to 
often employed calculations with analytical production cross sections.
This means that the flux of antiprotons in CR in reality may be 
\emph{lower} at the top of the atmosphere by at least 25-30\%.
If the latter is true, the reacceleration model (even without LB)
could still be the best one to describe propagation of nucleon species 
in the Galaxy. 
The inclusion of all known effects such as, e.g., sub-threshold
antiproton production on the abundant atmospheric N and O, 
may be important for evaluation of the correct atmospheric background.

We note that \citet{donato} claim to have obtained agreement with antiproton
measurements in a reacceleration plus convection model
using the parameters derived from B/C and sub-Fe/Fe ratios \citep{maurin}.
Apart from having one more free parameter (convection \emph{plus} reacceleration),
they in fact fitted B/C and sub-Fe/Fe ratios only at high energies
(since in their fitting procedure the high energy data overweights
few low energy points).
Their calculated ratios at low energies are higher than the Voyager and ACE data by
approximately 20\% or about $6\sigma$ \citep[see their Figs.\ 3, 4 in][]{maurin}. 
This is, however, where most of the problem lies. 
Besides, in their nuclear reaction treatment, they use the semiempirical cross sections by
\citeauthor{W-code}, which are not particulary accurate at low energies
while reacceleration models are sensitive to low energy behaviour of the cross 
sections.

\smallskip 
\section{CONCLUSION}\label{sec:conclusion}
In a previous paper we have shown that new more accurate measurements 
of the CR antiproton flux pose a challenge to existing CR propagation models.  
In particular, the antiproton flux and B/C ratio appear to
be inconsistent with measurements when computed in standard 
diffusion/reaccelation models.
In this paper we have demonstrated that this discrepancy can be resolved if
some part of the CR that we measure near the earth consists of a 
``fresh'' component accelerated in the LB.
The independent evidence for SN activity in the solar vicinity
in the last few Myr supports this idea.

Combining the measurements of the antiproton flux \emph{and}
B/C ratio to fix the diffusion coefficient, 
we have been able to construct a model consistent with 
measurements of important nuclei ratios 
in CR and derive elemental abundances in the LB.
Calculated isotopic abundance distributions of Be and B 
are in perfect agreement with CR data.
The abundances of three
radioactive isotopes in CR, which are often
used as ``radioactive clocks'' to determine the Galactic halo size,  
\e{10}{}Be, \e{26}{}Al, \e{36}{}Cl
are all consistent and indicate a halo size $z_h\sim4$ kpc based on 
the most accurate data by ACE spacecraft.
\e{54}{}Mn indicated a smaller halo, but this may be related
to its half-life uncertainty and/or cross section errors.
The derived fraction of LB component in CR is small compared to
Galactic CR and has a steep spectrum with a cutoff above 
several hundred MeV/nucleon.
Other experimental data (except maybe the overabundance
of Sc, Ti, V) do not contradict this hypothesis. 
The derived source overabundance of sub-Fe elements in the LB 
may be caused by trivial uncertainties in the production cross sections,
or could in principle arise from composition and/or evolution differences between 
the ISM in the LB and solar or Galactic average ISM. 
(LB may have evolved during the sun's lifetime of 4.5 Gyr.) 
This is suggested by 
the fact that the derived relative abundances of secondary elements in the LB
sources are systematically larger than in the Galactic
CR sources \citep[see, e.g., derivation of interstellar 
\e{12}{}C/\e{13}{}C ratio,][]{savage}.

The production cross sections if measured accurately
would help to distinguish between the different hypotheses; as of now, 
many important channels are not known accurately enough.
Such cross section errors lead to errors in important isotopic ratios, which, in turn,
are translated into  errors in propagation parameters.
In our treatment of Be and B production
cross sections, as well as some isotopes of other elements, 
we use all available data and our own fits to them, which
should be more accurate than semiempirical systematics by 
\citeauthor{W-code} and \citeauthor{ST-code}

\acknowledgements
The authors are grateful to M.\ Wiedenbeck for providing the ACE
isotopic abundances. I.\ V.\ M.\ is grateful to the Gamma Ray Group
of the Max-Planck-Institut f\"ur extraterrestrische Physik, 
where a part of this work has been done, for hospitality.
I.\ V.\ M.\ and S.\ G.\ M.\ acknowledge partial support from a
NASA Astrophysics Theory Program grant.

\appendix
\section{PRODUCTION CROSS SECTIONS OF ISOTOPES OF CARBON AND NITROGEN}\label{cs}

The production cross section of the most abundant $^{12}$C
isotope is one of the most poorly known.
All the data available to us on its production by $^{16}$O and
nitrogen isotopes on protons are summarized in Table~\ref{table4}.
The data include also production of $^{12}$B, and $^{12}$N, 
which decay to $^{12}$C with branching ratio 0.98, and
$^{13}$O with branching ratio 0.12.

\placetable{table4}
\placetable{table5}

Data on the production of $^{13}$C are somewhat more extensive,
but some important channels are not measured accurately
enough. 
Most of the data available are summarized in Table~\ref{table5}.
The data include also production of $^{13}$N, 
which decays to $^{12}$C with branching ratio 1. The production cross section
of $^{13}$B is very small, fractions of a mb.
Fortunately there are data on production of $^{13}$N by protons
on natural samples of oxygen and nitrogen \citep[see compilation by][]{t16lib}, 
which contain mostly $^{16}$O and $^{14}$N isotopes respectively.
$^{13}$N production cross section data probably include also production of $^{13}$O, but the
latter cross section must be very small (0.17 mb at 2100 MeV/nucleon).

The production cross section of the most abundant $^{14}$N
isotope is also poorly known.
All the data available to us are summarized in Table~\ref{table6}.
The compiled data
include also production of $^{14}$C, and $^{14}$O, 
which decay to $^{14}$N with branching ratio 1.

\placetable{table6}
\placetable{table7}

The main contributor to the production cross section of 
$^{15}$N is $^{16}$O (Table~\ref{table7}). 
The direct and indirect (via $^{15}$O)
production cross sections are almost equal.
The channel $p+^{16}$O $\to^{15}$O is well studied since
there is a large amount of data obtained on natural sample of oxygen 
\citep[see compilation by][]{t16lib}.

\begin{figure*}[!th]
\epsscale{\fsizeone}
\plottwo{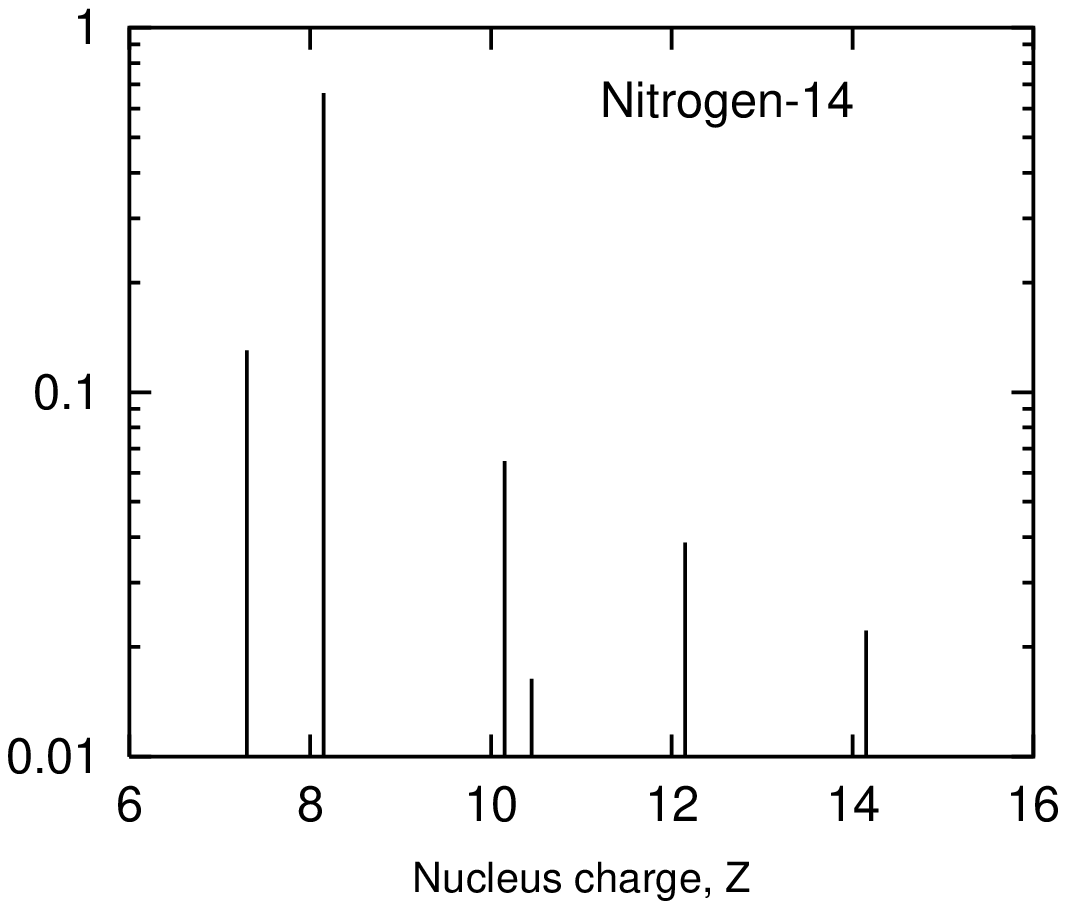}{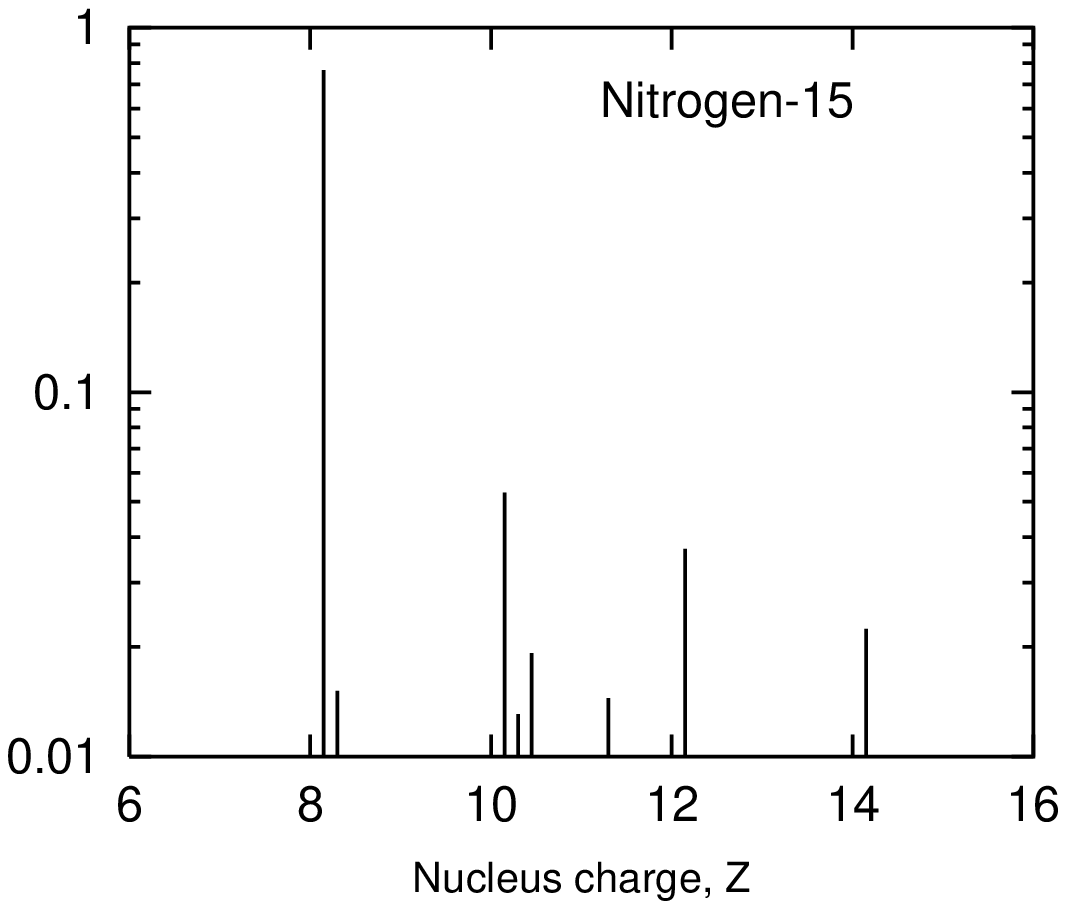}
\caption[f15a.eps,f15b.eps,f15c.eps,f15d.eps]{
Relative contributions of heavier isotopes to production of carbon and 
nitrogen isotopes are shown.
These contributions are determined from cumulative cross sections
of carbon and nitrogen isotopes weighted with 
the flux of corresponding primary isotope in CR at 1 GeV/nucleon.
The contributions of isotopes of a given element are indicated 
by separate lines. 
The scale gives the fraction of the C and N isotope produced by
a given primary isotope.
\label{fig:cn}}
\end{figure*}

The cumulative (sum over all channels) production cross sections
of carbon and nitrogen isotopes
multiplied by the flux of the corresponding primary isotope in CR at 1 GeV/nucleon
are shown in Fig.~\ref{fig:cn}.
The main contributor to the production of secondary carbon and nitrogen
is $^{16}$O, accounting for about 80\% in case of nitrogen isotopes.
However, in the case of carbon, disintegration of $^{16}$O gives only about 50\%
with an essential contribution from nitrogen isotopes (and $^{13}$C in
case of $^{12}$C). 

\placefigure{fig:cn}

The contribution of $^{15}$N to the production of $^{13}$C is especially
large (Table~\ref{table5}). This is based on only one experimental point
which seems too large compared to the production cross sections
on $^{14}$N and $^{16}$O. This indicates that the reason 
for the large fraction of $^{13}$C in calculated CR abundances 
compared to the measurements (see Sections \ref{sec:spectrum} and 
\ref{sec:results}) may
be errors in the cross sections.

There are more examples of discrepancies in $^{13}$C and $^{14}$N
production cross sections (Tables \ref{table5}, \ref{table6}, shown in bold font). 
The production cross section of $^{13}$C by $^{22}$Ne at 580 MeV/nucleon
differs significantly from that at 400 MeV/nucleon. 
The cross section of $^{13}$C by $^{26}$Mg measured by the same 
group at 370 and 576 MeV/nucleon differs by a factor of 4
(6.3 mb vs.\ 25 mb). A similar situation occurs with $^{14}$C
production by $^{22}$Ne at 400, 580, and 894 MeV/nucleon,
and by $^{26}$Mg at 371 and 576 MeV/nucleon (3.5 mb vs.\ 9 mb).
Fortunately, these latter cross sections do not contribute much
to production of $^{13}$C and $^{14}$N in CR, but these discrepancies
indicate the degree of overall uncertainty in the
production cross sections.

Cross section errors in production of carbon may lead to errors in the 
B/C ratio, which, in turn,
are translated into errors on the propagation parameters.
Because the CR measurements are now rather accurate, the errors
in the cross sections may cause many standard deviations when comparing
the model calculations with CR data.

It is clear that a more systematic approach to calculated cross sections
is required such as using, e.g., evaluated cross sections in future similar
work instead of only scarce experimental data or calculations by
stand-alone nuclear reaction models or phenomenological systematics. 
Such evaluated data files \citep{t16lib} have proved to be useful,
e.g., to study the production of radioisotopes for
medical and industrial applications using high power accelerators
\citep*{riper}.
At present, neither available experimental data nor any
of the current models or phenomenological systematics
can be used alone to produce a reliable evaluated activation cross section
library covering a wide range of target nuclides and incident
energies. Instead, such an evaluated library may be created 
by constructing excitation
functions using all available experimental data along with calculations
employing some of the most reliable codes in 
the regions of targets and incident energies where they are most applicable.
When there are reliable experimental data, they, rather than model results,
should be taken as the highest priority for the evaluation.
The development of such evaluated data libraries for astrophysical
applications is planned in the near future.

\placetable{table4}

\begin{table*}[!th]
\tablecolumns{6}
\tablewidth{0mm}
\begin{minipage}[tl]{0.67\textwidth}
\tablecaption{Collection of $^{12}$C production cross section data on protons.
\label{table4}}
\end{minipage} 

\begin{tabular}{cccccc}

\tablehead{
\colhead{Primary} &\colhead{Secondary} &\colhead{Energy,} &
\colhead{Cross section,} &\colhead{Error,} &\colhead{} \\
\colhead{Nucleus} &\colhead{Nucleus} &\colhead{MeV/nucleon} &
\colhead{mbarn} &\colhead{mbarn} &\colhead{Reference}
}
\startdata
 $^{14}$N  & $^{12}$C & \phn 377 &      56.90 & 0.05\tablenotemark{a} & 1  \\ 
 $^{14}$N  & ---      & \phn 516 &      52.10 & 0.05\tablenotemark{b} & 2  \\ 
 $^{15}$N  & ---      & \phn 373 &      30.00 & 0.05\tablenotemark{a} & 1  \\ 
 $^{16}$O  & ---      & \phn 389 &      33.90 & 0.05\tablenotemark{a} & 1  \\ 
 $^{16}$O  & ---      & \phn 516 &      33.60 & 0.05\tablenotemark{b} & 2  \\ 
 $^{16}$O  & ---      &     2100 &      32.30 &  4.80 & 3\medskip\\ 

 $^{16}$O  & $^{12}$B & \phn 516 &  \phn 1.10 & 0.30\tablenotemark{b} & 2  \\ 
 $^{16}$O  & ---      &     2100 &  \phn 1.45 & 0.17 & 3\medskip\\ 

 $^{14}$N  & $^{12}$N & \phn 516 &  \phn 1.10 & 0.30\tablenotemark{b} & 2  \\ 
 $^{16}$O  & ---      & \phn 516 &  \phn 0.30 & 0.30\tablenotemark{b} & 2  \\ 
 $^{16}$O  & ---      &     2100 &  \phn 0.40 & 0.07 & 3  \\ 
\enddata
\footnotesize
\tablenotetext{a}{Relative error.}
\tablenotetext{b}{Relative error is shown as indicated in \citet{We98}.}
\tablerefs{(1) \citet{We98}; (2) \citet{105}; (3) \citet{Ol83}.}
\end{table*}

\placetable{table5}


\begin{table*}[p]
\tablecolumns{6}
\tablewidth{0mm}
\begin{minipage}[tl]{0.67\textwidth}
\tablecaption{Collection of $^{13}$C production cross section data on protons.
\label{table5}}
\end{minipage} 

\begin{tabular}{cccccc}

\tablehead{
\colhead{Primary} &\colhead{Secondary} &\colhead{Energy,} &
\colhead{Cross section,} &\colhead{Error,} &\colhead{} \\
\colhead{Nucleus} &\colhead{Nucleus} &\colhead{MeV/nucleon} &
\colhead{mbarn} &\colhead{mbarn} &\colhead{Reference}
}
\startdata
 $^{14}$N    & $^{13}$C & \phn 377 &  \phn 7.60 & 0.05\tablenotemark{a} & 1  \\ 
 $^{14}$N    & ---      & \phn 516 &  \phn 9.60 & 0.05\tablenotemark{b} & 2  \\ 
 $^{15}$N    & ---      & \phn 373 & \bf  35.30 & 0.05\tablenotemark{a} & 1  \\ 
 $^{16}$O    & ---      & \phn 389 &      17.40 & 0.05\tablenotemark{a} & 1  \\ 
 $^{16}$O    & ---      & \phn 516 &      18.00 & 0.05\tablenotemark{b} & 2  \\ 
 $^{16}$O    & ---      &     2100 &      17.80 & 1.70 & 3  \\ 
 $^{20}$Ne   & ---      & \phn 414 &      15.30 & 0.05\tablenotemark{a} & 1  \\ 
 $^{20}$Ne   & ---      & \phn 534 &      15.70 & 0.05\tablenotemark{b} & 2  \\ 
 $^{22}$Ne   & ---      & \phn 377 &      17.80 & 1.40 & 4  \\ 
 $^{22}$Ne   & ---      & \phn 401 & \bf  15.30 & 0.05\tablenotemark{a} & 1  \\ 
 $^{22}$Ne   & ---      & \phn 581 & \bf  21.90 & 1.90 & 4  \\ 
 $^{22}$Ne   & ---      & \phn 894 &      19.00 & 1.60 & 4  \\ 
 $^{26}$Mg   & ---      & \phn 371 & \bf \phn 6.30 & 1.20 & 4  \\ 
 $^{26}$Mg   & ---      & \phn 576 & \bf  25.00 & 2.80 & 4\medskip\\ 

\e{nat}{}N($^{14}$N)& $^{13}$N & \multicolumn{3}{c}{multi-data} & 5  \\ 
 $^{14}$N    & ---      & \phn 377 &  \phn 7.40 & 0.10\tablenotemark{a} & 1  \\ 
 $^{14}$N    & ---      & \phn 516 &  \phn 7.50 & 0.10\tablenotemark{b} & 2  \\ 
 $^{15}$N    & ---      & \phn 373 &  \phn 4.30 & 0.30\tablenotemark{a} & 1  \\ 
\e{nat}{}O($^{16}$O)& ---      & \multicolumn{3}{c}{multi-data} & 5  \\ 
 $^{16}$O    & ---      & \phn 389 &  \phn 4.60 & 0.20\tablenotemark{a} & 1  \\ 
 $^{16}$O    & ---      & \phn 516 &  \phn 5.70 & 0.20\tablenotemark{b} & 2  \\ 
 $^{16}$O    & ---      &     2100 &  \phn 4.49 & 0.46 & 3  \\ 
 $^{20}$Ne   & ---      & \phn 414 &  \phn 5.10 & 0.20\tablenotemark{a} & 1  \\ 
 $^{20}$Ne   & ---      & \phn 534 &  \phn 4.10 & 0.20\tablenotemark{b} & 2  \\ 
 $^{22}$Ne   & ---      & \phn 377 &  \phn 0.50 & 0.10 & 4  \\ 
 $^{22}$Ne   & ---      & \phn 401 &  \phn 1.60 & 0.20\tablenotemark{a} & 1  \\ 
 $^{22}$Ne   & ---      & \phn 581 &  \phn 0.50 & 0.10 & 4  \\ 
 $^{22}$Ne   & ---      & \phn 894 &  \phn 0.70 & 0.20 & 4  \\ 
 $^{24}$Mg   & ---      & \phn 610 &  \phn 6.00 & 0.20\tablenotemark{b} & 2  \\ 
 $^{26}$Mg   & ---      & \phn 371 &  \phn 0.30 & 0.10 & 4  \\ 
 $^{26}$Mg   & ---      & \phn 576 &  \phn 0.10 & 0.10 & 4  \\ 
\enddata
\footnotesize
\tablecomments{Discrepancy in the data is shown in bold.}
\tablenotetext{a}{Relative error.}
\tablenotetext{b}{Relative error is shown as indicated in \citet{We98}.}
\tablerefs{(1) \citet{We98}; (2) \citet{105}; (3) \citet{Ol83}; (4) \citet{108}; 
(5) a compilation by \citet{t16lib}.}
\end{table*}

\placetable{table6}


\begin{table*}[p]
\tablecolumns{6}
\tablewidth{0mm}
\begin{minipage}[tl]{0.67\textwidth}
\tablecaption{Collection of $^{14}$N production cross section data on protons.
\label{table6}}
\end{minipage} 

\begin{tabular}{cccccc}

\tablehead{
\colhead{Primary} &\colhead{Secondary} &\colhead{Energy,} &
\colhead{Cross section,} &\colhead{Error,} &\colhead{} \\
\colhead{Nucleus} &\colhead{Nucleus} &\colhead{MeV/nucleon} &
\colhead{mbarn} &\colhead{mbarn} &\colhead{Reference}
}
\startdata
 $^{15}$N   & $^{14}$N &  \phn 373 &     27.60 & 0.05\tablenotemark{a} & 1  \\ 
 $^{16}$O   & ---      &  \phn 389 &     31.10 & 0.05\tablenotemark{a} & 1  \\ 
 $^{16}$O   & ---      &  \phn 516 &     31.00 & 0.05\tablenotemark{b} & 2  \\ 
 $^{16}$O   & ---      &      2100 &     31.00 & 3.30 & 3  \\ 
 $^{20}$Ne  & ---      &  \phn 414 &     25.80 & 0.05\tablenotemark{a} & 1  \\ 
 $^{22}$Ne  & ---      &  \phn 401 &     11.60 & 0.10\tablenotemark{a} & 1\medskip\\ 

 $^{15}$N   & $^{14}$C &  \phn 373 &     10.30 & 0.05\tablenotemark{a} & 1  \\ 
 $^{16}$O   & ---      &  \phn 389 & \phn 1.70 & 0.10\tablenotemark{a} & 1  \\ 
 $^{16}$O   & ---      &  \phn 516 & \phn 1.70 & 0.10\tablenotemark{b} & 2  \\ 
 $^{16}$O   & ---      &      2100 & \phn 3.69 & 0.38 & 3  \\ 
 $^{20}$Ne  & ---      &  \phn 414 & \phn 2.20 & 0.10\tablenotemark{a} & 1  \\ 
 $^{20}$Ne  & ---      &  \phn 534 & \phn 2.30 & 0.10\tablenotemark{b} & 2  \\ 
 $^{22}$Ne  & ---      &  \phn 377 & \phn 8.10 &  0.70 & 4  \\ 
 $^{22}$Ne  & ---      &  \phn 401 & \phn 7.70 & 0.05\tablenotemark{a} & 1  \\ 
 $^{22}$Ne  & ---      &  \phn 581 &     10.20 &  1.20 & 4  \\ 
 $^{22}$Ne  & ---      &  \phn 894 & \phn 8.60 &  0.90 & 4  \\ 
 $^{26}$Mg  & ---      &  \phn 371 & \bf \phn 3.50 &  0.70 & 4  \\ 
 $^{26}$Mg  & ---      &  \phn 576 & \bf \phn 9.00 &  1.30 & 4\medskip\\ 

 $^{16}$O   & $^{14}$O &  \phn 389 & \phn 1.30 & 0.30\tablenotemark{a} & 1  \\ 
 $^{16}$O   & ---      &  \phn 516 & \phn 1.20 & 0.30\tablenotemark{b} & 2  \\ 
 $^{16}$O   & ---      &      2100 & \phn 0.75 & 0.12 & 3  \\ 
 $^{20}$Ne  & ---      &  \phn 534 & \phn 1.00 & 0.30\tablenotemark{b} & 2  \\ 
 $^{24}$Mg  & ---      &  \phn 610 & \phn 1.50 & 0.30\tablenotemark{b} & 2  \\ 
\enddata
\footnotesize
\tablecomments{Discrepancy in the data is shown in bold.}
\tablenotetext{a}{Relative error.}
\tablenotetext{b}{Relative error is shown as indicated in \citet{We98}.}
\tablerefs{(1) \citet{We98}; (2) \citet{105}; (3) \citet{Ol83}; (4) \citet{108}.}
\end{table*}

\placetable{table7}


\begin{table*}[p]
\tablecolumns{6}
\tablewidth{0mm}
\begin{minipage}[tl]{0.67\textwidth}
\tablecaption{Collection of $^{15}$N production cross section data on protons.
\label{table7}}
\end{minipage} 

\begin{tabular}{cccccc}

\tablehead{
\colhead{Primary} &\colhead{Secondary} &\colhead{Energy,} &
\colhead{Cross section,} &\colhead{Error,} &\colhead{} \\
\colhead{Nucleus} &\colhead{Nucleus} &\colhead{MeV/nucleon} &
\colhead{mbarn} &\colhead{mbarn} &\colhead{Reference}
}
\startdata
 $^{16}$O    & $^{15}$N & \phn 389 &      33.60 & 0.05\tablenotemark{a} & 1  \\ 
 $^{16}$O    & ---      & \phn 516 &      34.90 & 0.05\tablenotemark{b} & 3  \\ 
 $^{16}$O    & ---      &     2100 &      34.30 & 3.30 & 4  \\ 
 $^{20}$Ne   & ---      & \phn 414 &      24.00 & 0.05\tablenotemark{a} & 1  \\ 
 $^{20}$Ne   & ---      & \phn 534 &      27.80 & 0.05\tablenotemark{b} & 3  \\ 
 $^{22}$Ne   & ---      & \phn 377 &      36.20 & 2.10 & 5  \\ 
 $^{22}$Ne   & ---      & \phn 401 &      32.90 & 0.05\tablenotemark{a} & 1  \\ 
 $^{22}$Ne   & ---      & \phn 581 &      39.00 & 2.50 & 5  \\ 
 $^{22}$Ne   & ---      & \phn 894 &      33.50 & 2.10 & 5  \\ 
 $^{24}$Mg   & ---      & \phn 610 &      14.00 & 0.10\tablenotemark{b} & 3  \\ 
 $^{26}$Mg   & ---      & \phn 371 & \bf  19.70 & 2.10 & 5  \\ 
 $^{26}$Mg   & ---      & \phn 576 & \bf  29.90 & 3.00 & 5\medskip\\ 
 
 $^{22}$Ne   & $^{15}$C & \phn 377 &  \phn 0.70 &  0.20 & 5  \\ 
 $^{22}$Ne   & ---      & \phn 581 &  \phn 0.80 &  0.20 & 5  \\ 
 $^{22}$Ne   & ---      & \phn 894 &  \phn 0.60 &  0.10 & 5  \\ 
 $^{26}$Mg   & ---      & \phn 371 &  \phn 0.90 &  0.30 & 5  \\ 
 $^{26}$Mg   & ---      & \phn 576 &  \phn 0.40 &  0.30 & 5\medskip\\ 

\e{nat}{}O($^{16}$O)  & $^{15}$O & \multicolumn{3}{c}{multi-data} & 6  \\ 
 $^{16}$O    & ---      & \phn 389 &      30.70 & 0.05\tablenotemark{a} & 1  \\ 
 $^{16}$O    & ---      & \phn 516 &      30.30 & 0.05\tablenotemark{b} & 3  \\ 
 $^{16}$O    & ---      &     2100 &      27.30 & 2.60 & 4  \\ 
 $^{20}$Ne   & ---      & \phn 414 &      14.90 & 0.10\tablenotemark{a} & 1  \\ 
 $^{20}$Ne   & ---      & \phn 534 &      16.20 & 0.10\tablenotemark{b} & 3  \\ 
 $^{21}$Ne   & ---      & \phn 520 &  \phn 7.80 & 0.30\tablenotemark{a} & 2  \\ 
 $^{22}$Ne   & ---      & \phn 377 &  \phn 2.00 & 0.30 & 5  \\ 
 $^{22}$Ne   & ---      & \phn 401 &  \phn 1.60 & 0.20\tablenotemark{a} & 1  \\ 
 $^{22}$Ne   & ---      & \phn 581 &  \phn 1.60 & 0.30 & 5  \\ 
 $^{22}$Ne   & ---      & \phn 894 &  \phn 2.80 & 0.40 & 5  \\ 
 $^{22}$Na   & ---      & \phn 520 &      10.90 & 0.30\tablenotemark{a} & 2  \\ 
 $^{23}$Na   & ---      & \phn 517 &      11.10 & 0.30\tablenotemark{a} & 2  \\ 
 $^{24}$Mg   & ---      & \phn 610 &  \phn 8.60 & 0.10\tablenotemark{b} & 3  \\ 
 $^{25}$Mg   & ---      & \phn 514 &  \phn 6.00 & 0.30\tablenotemark{a} & 2  \\ 
 $^{26}$Mg   & ---      & \phn 371 &  \phn 1.10 & 0.30 & 5  \\ 
 $^{26}$Mg   & ---      & \phn 576 &  \phn 0.20 & 0.20 & 5  \\ 
 $^{26}$Al   & ---      & \phn 508 &  \phn 4.10 & 0.30\tablenotemark{a} & 2  \\ 
 $^{27}$Al   & ---      & \phn 511 &  \phn 7.10 & 0.20\tablenotemark{a} & 2  \\ 
 $^{28}$Si   & ---      & \phn 506 &  \phn 5.60 & 0.12\tablenotemark{a} & 2  \\ 
 $^{29}$Si   & ---      & \phn 508 &  \phn 2.40 & 0.30\tablenotemark{a} & 2  \\ 
\enddata
\footnotesize
\tablecomments{Discrepancy in the data is shown in bold.}
\tablenotetext{a}{Relative error.}
\tablenotetext{b}{Relative error is shown as indicated in \citet{We98}.}
\tablerefs{(1) \citet{We98}; (2) \citet{We98prc}; (3) \citet{105}; (4) \citet{Ol83}; (5) \citet{108}; 
(6) a compilation by \citet{t16lib}.}
\end{table*}

\placefigure{fig:cn}
\vspace*{20mm}
\begin{figure*}[!th] 
\epsscale{\fsizeone}
\plottwo{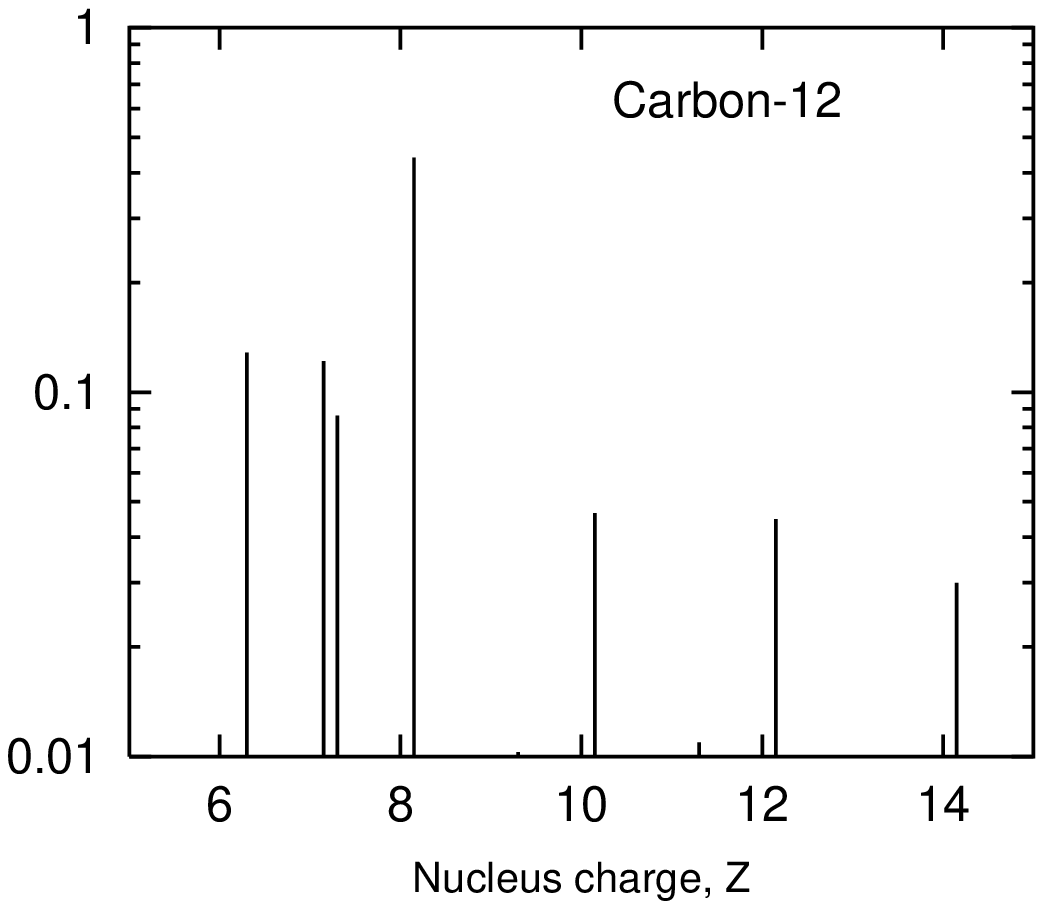}{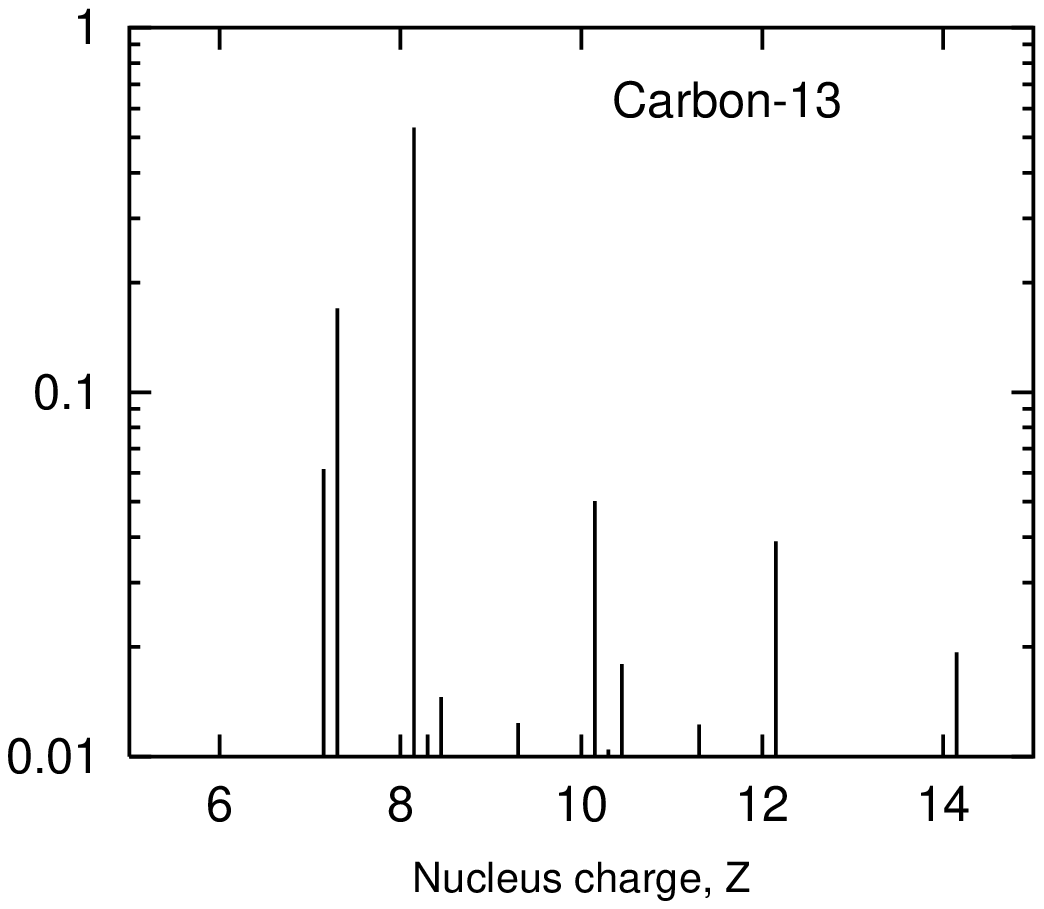}
\vspace{-40mm}
\end{figure*}

\end{document}